%% file: kortenka_preprint.tex
\documentclass[preprint]{aastex}

\usepackage{epsfig,graphics,ifthen}

\newcommand{\etal}{{\it et al.\ }}
\newcommand{\msc}[1]{{\mbox{\scriptsize{#1}}}}
\newcommand{\msct}[1]{{\mbox{\tiny{#1}}}}

\newcommand{\mar}[1]{\marginpar{\vspace{15pt}\textsf{\footnotesize #1}}}

\renewcommand{\L}[1]{L\ensuremath{_\msc{#1}}}

\newcommand{\fun}[1]{\ensuremath{f^\msct{UN}_\msct{#1}}}
\newcommand{\fnt}[1]{\ensuremath{f^\msct{NT}_\msct{#1}}}

\newcounter{mypage}

\newcommand{\mylabel}[1]{
        \setcounter{mypage}{\value{page}}
        \addtocounter{mypage}{-1}
        \refstepcounter{mypage}
        \label{#1}}

\begin{document}

\title{\textmd{\small \textsl{Icarus} manuscript \#I08618, submitted 16 May 2003}\\[1cm]
Survival of Trojan-Type Companions of Neptune\\
During Primordial Planet Migration} 

\author{
Stephen J. Kortenkamp$^{\ast,\dag}$, 
Renu Malhotra$^{\ast}$, and 
Tatiana Michtchenko$^{\ddag}$}

\normalsize

\section*{\centering {\normalsize ABSTRACT}}

\noindent
We investigate the survivability of Trojan-type companions of Neptune
during primordial radial migration of the giant planets Jupiter,
Saturn, Uranus, and Neptune.  We adopt the usual planet migration
model in which the migration speed decreases exponentially with a
characteristic time scale $\tau$ (the e-folding time).  We perform a
series of numerical simulations, each involving the migrating giant
planets plus $\sim$1000 test particle Neptune Trojans with initial
distributions of orbital eccentricity, inclination, and libration
amplitude similar to those of the known jovian Trojans asteroids.  We
analyze these simulations to measure the survivability of Neptune's
Trojans as a function of migration rate.  We find that orbital
migration  with the characteristic time scale $\tau = 10^6$ years
allows about 35\% of pre-existing Neptune Trojans to survive to
$5\tau$, by which time the giant planets have essentially reached
their final orbits.  In contrast, slower migration with $\tau = 10^7$
years yields only a $\sim$5\% probability of Neptune Trojans surviving
to a time of $5\tau$.  Interestingly, we find that the loss of Neptune
Trojans during planetary migration is not a random diffusion process.
Rather, losses occur almost exclusively during discrete episodes when
Trojan particles are swept by secondary resonances associated with
mean-motion commensurabilities of Uranus with Neptune.  These
secondary resonances arise when the circulation frequencies, $f$, of
critical arguments for Uranus-Neptune mean-motion near-resonances
(e.g., \fun{1:2}, \fun{4:7}) are commensurate with harmonics of the
libration frequency of the critical argument for the Neptune-Trojan
1:1 mean-motion resonance (\fnt{1:1}).  Trojans trapped in the
secondary resonances typically have their libration amplitudes 
amplified until they escape the 1:1 resonance with Neptune.  Trojans
with large libration amplitudes are susceptible to loss during
sweeping by numerous high order secondary resonances (e.g.,
$\fun{1:2}\,\approx\,11\fnt{1:1}$). However, for the slower migration, with
$\tau = 10^7$ years, even tightly bound Neptune Trojans with libration
amplitudes below $10^\circ$ can be lost when they become trapped in
1:3 or 1:2 secondary resonances between \fun{1:2} and \fnt{1:1}.  With
$\tau = 10^7$ years the 1:2 secondary resonance was responsible for
the single greatest episode of loss, ejecting nearly 75\% of existing
Neptune Trojans.  This episode occurred during the late stages of
planetary migration when the remnant planetesimal disk would have been
largely dissipated.  We speculate that if the number of bodies
liberated during this event was sufficiently high they could have
caused a spike in the impact rate throughout the solar system.

\vspace{0.5cm}

\noindent $\ast$ University of Arizona, Tucson, AZ

\noindent $\dag$ Planetary Science Institute, Tucson, AZ

\noindent $\ddag$ Universidade de S\~ao Paulo, S\~ao Paulo, Brasil

\newpage


\vspace{-0.5cm}
\section*{\centering \normalsize I. INTRODUCTION}
\vspace{-0.2cm}

Nearly 1500 asteroids are known to share Jupiter's orbit, locked in
1:1 mean-motion resonance with the giant planet.  These so-called
Trojan asteroids form two swarms in the Lagrange equilibrium regions
leading (\L{4}) and trailing (\L{5}) Jupiter in its orbit by
$60^\circ$.  Only a single non-jovian Trojan-type companion is yet
known in the outer solar system, minor planet 2001\,QR322 in Neptune's
trailing \L{5} region (Chiang 2003; Marsden 2003).  The lack of
saturnian Trojans may be explained by the chaotic effects of
resonances with other planets, which disrupt what would otherwise be
stable Trojan regions (Holman \& Wisdom 1993; de la Barre \etal 1996).
Trojans of Uranus may also have been largely depleted by dynamical
instabilities over the age of the solar system (Nesvorn\'y \& Dones
2002).  The paucity of Neptune Trojans, however, is more puzzling.
Most investigations into the stability of hypothetical Neptune Trojans
(e.g., Weissman \& Levison 1997; Nesvorn\'y \& Dones 2002) suggest
that if a primordial population once existed then some sizable remnant
($\sim$50\%) of it should be preserved today.  Prior to 2001 there had
been no  discoveries of Neptune Trojans, neither serendipitously nor
in dedicated surveys (Chen \etal 1997).  This fact was used to set
crude limits on the size distribution of any existing population
(Nesvorn\'y \& Dones 2002).  However, these upper limits are weak,
and do not preclude a population of Neptune Trojans  that may exceed
the population of Jupiter Trojans.  Nevertheless, the current apparent
scarcity of Neptune Trojans is interesting from a dynamical point of
view.  This, together with the orbital characteristics of 2001\,QR322,
may provide important clues to Neptune's origin and the primordial
orbital evolution of the giant planets.

The early dynamical evolution of the outer solar system is thought to
involve significant radial migration of the four giant planets
(Fern\'andez \& Ip 1984, 1996; Hahn \& Malhotra 1999).  The cause of
this migration is the planets' dynamical clearing of the residual
planetesimal disk during the late stages of planet formation.  Some
clues that suggest past planetary migration include the existence of
the Oort cloud of comets, orbital characteristics of plutinos and
other resonant Kuiper belt objects (KBOs) trapped in mean-motion
resonance with Neptune (Malhotra 1993, 1995; Chiang \& Jordan 2002),
the high inclination classical KBOs (Gomes 2003), and depletion of the
outer asteroid belt (Liou \& Malhotra 1997).  Detailed descriptions of
the migration process can be found in Fern\'andez \& Ip (1984, 1996)
and Hahn \& Malhotra (1999).  Here we simply re-state the general
scenario, highlighting the effects on Neptune's orbit.

In the beginning, Neptune is embedded in a planetesimal disk and we
will assume it has a mean specific orbital angular momentum roughly
equal to that of the planetesimals in its gravitational vicinity.  Of
those planetesimals encountered by Neptune about equal numbers will be
scattered inward and outward, thus initially there is no net change in
Neptune's orbital energy or angular momentum, so no migration of
Neptune's orbit.  Most planetesimals scattered outward eventually
return to Neptune's region, with only minor losses from those
receiving sufficient stellar perturbations to be boosted into the Oort
cloud.  These returning planetesimals can again be scattered by
Neptune either inward or outward.  Thus, most of the energy and
angular momentum lost by Neptune to outward scattering can be
recaptured upon return of the planetesimal.  However, those
planetesimals scattered inward by Neptune have the opportunity to
encounter Uranus, Saturn, and Jupiter.  Jupiter essentially controls
the overall dynamics because it is very effective at ejecting
planetesimals from the solar system.  Planetesimals scattered inward
by Neptune and subsequently ejected by Jupiter cannot return for
another encounter with Neptune (or Jupiter).  Thus, Neptune retains
the energy and angular momentum gained from them.  In this scenario
Jupiter provides more than the required energy and angular momentum to
clear the remnant planetesimal disk, with the extra going toward
expansion of Neptune's orbit.  Jupiter thus migrates inward, though
only slightly owing to its large mass.  The early evolution of Saturn
and Uranus are somewhat analogous to Neptune's, with each of these
planets also migrating outward at Jupiter's expense.

Modeling of the cause and effects of planetary migration by
planetesimal scattering (e.g., Hahn \& Malhotra 1999; Gomes 1997) has
yielded approximate semi-major axis migration distances, $\Delta a$,
of $-0.2$, 0.8, 3.0, and 7.0\,AU for Jupiter, Saturn, Uranus, and
Neptune, respectively.  The present semi-major axes of these planets
are about 5.2, 9.5, 19.2, and 30.1\,AU.  Effective constraints on the
early orbital evolution of the giant planets can be found in the
orbital eccentricities of Pluto and other Kuiper belt objects trapped
in 2:3 mean-motion resonance with Neptune.  Using these constraints,
Malhotra (1993, 1995) estimated that 5\,AU should be considered a
minimum migration distance for Neptune.  Gomes (1997, 2000) and
Malhotra (1998) suggested that a 9--10\,AU migration might be
necessary to account for the plutino inclinations.  Constraints on the
timescale of migration have come from the magnitude of $\Delta a$ for
Neptune (Hahn \& Malhotra 1999), from relative populations of the 2:3
and 1:2 resonant KBOs (Malhotra 1995; Friedland 2000; Chiang \& Jordan
2002), and plutino inclinations (Gomes 1997).  With the discovery of
2001 QR\,322, a new class of small body in resonance with Neptune has
become available for study.  Our aim with this paper is to begin
examining the constraints such bodies may place on the early dynamical
evolution of the giant planets.

The critical argument for a Trojan-type particle, an object in 1:1
mean-motion resonance with a planet, is given by  $\phi_\msct{1:1} =
\lambda - \lambda_\msct{pln}$, where $\lambda$ and
$\lambda_\msct{pln}$ are the mean longitudes of the particle and
planet.  When a particle and planet are not in resonance
$\phi_\msct{1:1}$ circulates through all angles 0 to $360^\circ$.  For
a particle trapped in the planet's leading \L{4} or trailing \L{5}
regions, $\phi_\msct{1:1}$ librates about $+60^\circ$ or $-60^\circ$,
respectively.  The full magnitude of the difference between the
extremal values of $\phi_\msct{1:1}$ is called the libration
amplitude.  The time required for one complete libration of
$\phi_\msct{1:1}$ is the libration period.  The effects of a planet's
radial migration on the libration amplitude of a Trojan-type companion
has been explored for the planar circular restricted three-body
problem consisting of the sun, planet, and massless Trojan particle.
Fleming and Hamilton (2000) derived the expression $A_f/A_i \propto
(a_f/a_i)^{-1/4}$, where $A$ is the Trojan libration amplitude, $a$ is
the planet's semi-major axis, and the subscripts denote the final and
initial values.  Inward radial migration increases $A$ while outward
migration decreases $A$.  Numerical simulations by Fleming and
Hamilton (2000) confirmed the validity of their expression for small
initial libration amplitudes ($A_i<30^\circ$) and ``slow'' (adiabatic)
migration.  This single-planet model of Trojan stability during
planetary migration is useful for demonstrating basic principles of
the effects of migration.  A straight forward application of this
expression to our migration distances implies a slight decrease of
about 5\% in the libration amplitude of Neptune's Trojans.  However,
we show here that the gravitational perturbations of multiple
migrating giant planets, especially at resonance crossings, have a
profound effect on Trojan orbits that dwarfs the simple adiabatic
response due to a single migrating planet.

The effects of radial migration of all four giant planets on
pre-existing Neptune Trojans have not received much attention.  As far
as we are aware, the only published study of the problem is that of
Gomes (1998).  Gomes used both linear and exponential migration  and
found significant survival of Neptune Trojans in $N$-body simulations
which had all four giant planets migrating.  Gomes included 100
initial Trojan-type test particles and generally migrated the planets
using the conventional $\Delta a$ values given above.  In linear
migration models with a $10^6$ year time scale Gomes found 82
particles survived as Neptune Trojans after migration.  For a time
scale of $10^7$ years 53 Trojan particles survived.  Using exponential
migration, Gomes found that 30 Neptune Trojans survived for $2\times
10^7$ years in a model with characteristic time scale $\tau = 1.5
\times 10^6$ years ($\tau$ is the e-folding time for migration) and
that 49 survived $2\times 10^7$ years with $\tau = 2.5 \times 10^6$
years, suggesting an increase in Trojan survivability with longer
migration time scales.  However, that trend was reversed in a model
with $\tau = 10^7$ years, where Gomes found only 30 Neptune Trojans
surviving for $5 \times 10^7$ years.

The relatively small number of initial test particles used by Gomes
(1998) and the inconsistent results for different migration rates
prevent a clear assessment of the general survivability of Neptune
Trojans during primordial radial migration of the four giant planets.
In addition, Gomes did not seek to identify the mechanism by which
Neptune Trojans are destabilized during planetary migration.  In the
present paper we have re-evaluated the problem using more than an
order of magnitude more particles.  Our simulations use only
exponential radial migration as this seems more realistic than linear
migration (Fern\'andez \& Ip 1984, 1996; Hahn \& Malhotra 1999).  We
find that there is a clear trend of decreasing Trojan survivability
with increasing migration time scale.  We also describe in detail the
primary mechanism responsible for destabilization and loss of Neptune
Trojans during migration.  The remainder of this paper is organized as
follows.  In Section II we describe our initial conditions, the
numerical model, and results from a series of simulations.  In Section
III we conclude by discussing some implications of our results in the
context of our current understanding of planet migration and earlier
work on Neptune Trojan stability.

\vspace{-0.5cm}
\section*{\centering {\normalsize II. METHOD AND RESULTS}} 
\vspace{-0.2cm}

\subsection*{\normalsize IIa. Initial Conditions and Numerical Model}

\mar{Tab.~\ref{tab:model_parameters}}

Unless otherwise noted, each of our simulations included the four
giant planets Jupiter, Saturn, Uranus, and Neptune.  The radial
migration of these planets followed a smooth time variation of their
semi-major axes, $a$.  A time scale $\tau$ was used to characterize
the migration, where $a(t) = a(0) + \Delta a [1 - \exp(-t/\tau)]$ and
$\Delta a$ is the desired amount of total migration at time $t =
\infty$.  Following Malhotra (1995), we adopt $\Delta a = -0.2$, 0.8,
3.0, and 7.0\,AU, respectively for Jupiter, Saturn, Uranus, and
Neptune.  Table~\ref{tab:model_parameters} lists the masses and
initial configuration of these planets for our simulations.  The
masses of the terrestrial planets were added to the mass of the sun.

Orbital evolution was followed using the Wisdom-Holman (1991)
symplectic integration technique with radial migration modifications
included as non-gravitational forces (see Cordeiro \etal 1997).  The
mutual gravitational perturbations of the planets were included
self-consistently even as their orbital spacing was expanding.  The
numerical simulation code is similar to that  used in earlier work
(Malhotra 1995).  For all simulations we used a time step of 6 months,
roughly $1/25$ the orbital period of the innermost planet (Jupiter).
This time step is sufficiently small to reliably integrate the orbits
of the four giant planets and the test particles over the time scales
we studied (Wisdom \& Holman 1992).

The initial conditions for the Neptune Trojans (modeled as massless
test particles) were obtained as follows.  We started with the
population of Jupiter's current Trojan companions as of January 2002,
a total of 1171 asteroids, having downloaded their orbital elements
from the Minor Planet Center database
(cfa-www.harvard.edu/iau/lists/JupiterTrojans.html).  These asteroids
are roughly evenly divided between Jupiter's leading \L{4} (57\%) and
trailing \L{5} (43\%) Lagrange equilibrium regions.  The orbital
elements of these asteroids were then assigned to test particles at
Neptune's initial orbit at 23\,AU, with appropriate transformations of
semi-major axes, arguments of pericenter, and longitudes of ascending
node.  Eccentricities and inclinations were not changed.  This
transformation was successful in placing most of the original
population in 1:1 mean-motion resonance with Neptune.  To verify this,
during the first 25,000 years of each simulation we monitored the
critical argument of each Neptune-Trojan pair,
$\phi^\msct{NT}_\msct{1:1} = \lambda - \lambda_\msct{Nep}$.    All
particles that maintained $0^\circ < |\phi^\msct{NT}_\msct{1:1}| <
180^\circ$ for the first 25,000 years were considered members of the
``initial'' resonant population.  This initial 25,000 years is about
four times the libration period of $\phi^\msc{NT}_\msc{1:1}$ for
Neptune's initial orbit at 23 AU\@.  Because of differences in the
parameter $\tau$, the number of particles in these initial populations
at 25,000 years varied slightly in different simulations, but was
always greater than 1000.  During the remainder of each simulation
$\phi^\msc{NT}_\msc{1:1}$ for each particle was calculated every 100
time steps (50 years) and the extremal values of
$\phi^\msc{NT}_\msc{1:1}$ were updated every $10^4$ years.

Our initial populations of Trojan particles (at {\sf T}$_\msc{o} =
25,000$ years) had a spread in semi-major axis $\Delta a \simeq 0.3$
AU, eccentricity $e \leq 0.2$, inclinations $i \leq 40^\circ$, and
libration amplitudes $A \leq 150^\circ$.  The range in ($a,e,i$) space
for these initial populations covers roughly the same volume as the
populations used by Nesvorn\'y and Dones (2002) for their study of
long term stability of Neptune Trojans subject to planetary
perturbations in the present configuration of the solar system.  They
found that roughly 50\% of the Neptune Trojan population survived
after $4\times 10^9$ years.  A direct application of their stability
results to our initial population is not possible because we used a
different tighter planetary configuration.  However, qualitatively
their results suggest that our initial conditions more than adequately
cover the most stable regions of the phase space for Neptune Trojans.

To increase run-time efficiency we removed particles from the
simulations if their libration amplitude exceeded $180^\circ$.  Thus,
Neptune could not regain Trojans after they had been lost.  A limited
number of simulations were carried out that tracked all particles for
the full duration of the simulation.  In these simulations a small
number of  particles on large tadpole orbits around either the leading
\L{4} or trailing \L{5} were found to be capable of transitioning to
and from resonant horseshoe orbits as the planets migrated.  However,
recapture of lost Trojans by this process was a short-term effect.
The counts of surviving Trojans in the primary runs typically differed
only by one or two particles compared to the runs that followed all
particles for the full duration of the simulation.

\subsection*{\normalsize IIb. Migration Simulations}

\mar{Fig.~\ref{fig:planets}}

Figure~\ref{fig:planets} shows the evolution of the semi-major axes of
the four giant planets as a function of time expressed in units of the
characteristic migration time scale $\tau$ up to a time of $8\tau$.
With this exponential migration model, radial migration of all planets
is 99.33\% complete after a time of $5\tau$.  By this time orbital
evolution of the planets (and of their Trojan companions) is dominated
by gravitational perturbations from the planets in their near-final
orbits rather than the effects of migration.  Because the final
configuration in each of our simulations cannot exactly match the
current configuration of the solar system, we confine our study of
Trojan survival to times less than or equal to $5\tau$.  For stability
analysis of Neptune Trojans beyond $5\tau$ we defer to gigayear
simulations that use the precise present orbital configuration of the
solar system (Weissman \& Levison 1997; Nesvorn\'y \& Dones 2002).
These studies have shown that if a post-migration population of
Neptune Trojans existed then some significant remnant ($\sim$50\%)
should have survived on a 4 billion year time scale.

\mar{Fig.~\ref{fig:tau_6}}

Figure~\ref{fig:tau_6} shows results from a relatively rapid migration
simulation with $\tau = 10^6$ years.  At the initial time {\sf
T}$_\msc{o} = 25,000$ years there were 1031 particles trapped in
Neptune's leading \L{4} and trailing \L{5} regions.  The top panel of
Fig.~\ref{fig:tau_6} shows a histogram of the extremal values of
$\phi^\msct{NT}_\msct{1:1}$ for this initial population.  The five
lower panels of Fig.~\ref{fig:tau_6} show similar histograms of the
Trojan population surviving after integer increments of the
characteristic migration time scale $\tau$.  After a time of $5\tau$
about 38\% of the initial Trojan population remained in 1:1 resonance
with Neptune.  This result is reasonably consistent with that of Gomes
(1998), who found a Neptune Trojan survival probability of 30-50\% for
$\tau = 1.5$ and $2.5 \times 10^6$ years.

\mar{Fig.~\ref{fig:tau_7}}

Figure~\ref{fig:tau_7} shows results from a simulation where the
migration rate was an order of magnitude slower than that of
Fig.~\ref{fig:tau_6}, with $\tau = 10^7$ years.  The initial
population at 25,000 years was nearly identical to the previous case,
with 1025 test particles trapped in the leading and trailing Lagrange
regions.  However, after a time of $5\tau$ most of the Trojans had
been lost, with only 1.5\% of the initial population surviving.  The
simulations shown in Figs.~\ref{fig:tau_6} and~\ref{fig:tau_7}
represent examples of our key result: survivability of Neptune Trojans
depends on the characteristic migration time scale.  Although the
simulation with $\tau = 10^7$ years ran for 10 times longer than that
with $\tau = 10^6$ years,  the greater losses in Fig.~\ref{fig:tau_7}
cannot be explained simply by the longer duration of the simulation.
In neither case were the losses the result of a gradual random
diffusive process.

\mar{Fig.~\ref{fig:tau_histogram}}

Because of the inherent chaotic nature of $N$-body simulations,
identical initial conditions modeled on different computers (different
processors and/or compilers) produced results that were generally in
agreement but differed in some details.  For example, some Trojans
lost after $5\tau$ in a $\tau = 10^7$ year simulation on a Sun Sparc
Ultra\,5 were survivors when the simulation was run on an Intel Xeon
CPU, and vice versa.  We ran simulations for each value of $\tau$ on
four different types of processors -- the Sparc, two Xeons (1.7 and
2.4 GHz), and a Digital Alpha\,EV6.  Figure~\ref{fig:tau_histogram}
shows a composite histogram of the survivability of Neptune Trojans
from runs on these four types of processors for simulations with $\tau
= 10^6, 2.5\times10^6, 5\times10^6$, and $10^7$ years.  These four
runs for each value of $\tau$ effectively quadruple the number of
initial test particles in each case, resulting in over 4000 initial
Trojan test particles for each value of $\tau$.
Figure~\ref{fig:tau_histogram} displays an obvious trend: Neptune
loses Trojans during migration and it loses more Trojans for longer
migration time scales.  Comparing the survival statistics in
Fig.~\ref{fig:tau_histogram} at equal durations in time (e.g., at
$5\times 10^6$ years for different values of $\tau$) reveals that
these losses cannot be explained by a constant diffusion with time.

\mar{Fig.~\ref{fig:jsn}}

In all simulations shown in Fig.~\ref{fig:tau_histogram} Neptune is
migrating from just exterior to the 3:5 mean-motion resonance with
Uranus to approximately its present orbit just interior to the 1:2
mean-motion resonance with Uranus (see Fig.~\ref{fig:planets}).
Between the 3:5 and 1:2, Neptune and its Trojans cross numerous other
higher order uranian resonances, such as the 4:7 and 5:9.  All of
these resonances play a role in destabilizing Neptune Trojans during
migration.  Simple confirmation of this was found by repeating some
simulations without Uranus (see Fig.~\ref{fig:jsn}).  In these
simulations nearly all initial Neptune Trojan particles survived to a
time of $5\tau$.

The importance of Uranus in this problem should not be entirely
unexpected.  In an ideal case Uranus is the planet that comes the
closest to, and strays the farthest from, Neptune's Trojans.  For a
Trojan initially trapped exactly at Neptune's \L{4} or \L{5} point
(zero libration amplitude), Uranus passes within 7\,AU at conjunction
and is as distant as 39\,AU at opposition.  Neptune, on the other
hand, remains at a distance of 23\,AU\@.  Accounting for the
difference in planetary mass, the gravitational force from Uranus
acting on the particle varies, from about nine times stronger than the
force from Neptune at conjunction to about one third weaker at
opposition.  These uranian perturbations will repeat with the synodic
period of Uranus and Neptune.  For a Trojan particle with non-zero
libration amplitude, its position will vary somewhat at each
conjunction with Uranus.  Thus, the uranian perturbations, while
roughly of the same magnitude, will at times accelerate and others
decelerate the particle's motion with respect to Neptune.  The net
effect over many libration periods amounts to destructive interference
which does not destabilize the Trojan particle's resonant
configuration with Neptune.  However, if the Uranus-Neptune-Trojan
orbital orientation becomes periodic in time (a three-body resonance)
then the uranian perturbations could buildup constructively, driving
the Trojan particle to larger or smaller libration amplitude.

Note that for the initial conditions in
Tab.~\ref{tab:model_parameters} the gravitational forces acting on
Neptune's Trojans from Saturn and Jupiter are always larger than the
force from Neptune.  Saturnian perturbations range from about 13 to 3
times those from Neptune, at conjunction and opposition with the
Trojan particle, respectively.  Jovian perturbations are stronger
still, 30 times neptunian at conjunction and 12 times at opposition.
The high retention of Trojans in simulations that did not include
Uranus (Fig.~\ref{fig:jsn}) demonstrate that the direct saturnian and
jovian perturbations do not dominate the stability of Neptune Trojans
during planetary migration.  Jupiter and Saturn have short synodic
periods with respect to Neptune (about 14 and 34 years).  Considering
the $\sim10^4$ year libration period of Neptune's Trojans, the high
frequency low amplitude perturbations from Saturn and Jupiter can be
considered in an orbit-averaged sense, as if the mass of these two
distant planets were distributed in a ring about their orbits.
Results shown in Fig.~\ref{fig:jsn} and the simple schematic described
above do not preclude destabilizing roles for Jupiter and Saturn
acting indirectly through perturbations on Uranus (Michtchenko \etal
2001).

\mar{Fig.~\ref{fig:lib_amp}}

Close scrutiny of the orbital evolution of some of the Trojans lost in
our simulations strengthens the case against Uranus acting alone.
Figure~\ref{fig:lib_amp} provides an example of the evolution of
Trojan libration amplitudes in a simulation with $\tau = 5\times10^6$
years.  In this simulation about 95\% of the initial population of
Neptune Trojans was lost by a time of $5\tau$.  Nearly all Trojans
that were lost between a time of $\tau$ and $5\tau$ evolved out of
resonance during a discrete number of prolonged episodes.  The top
panel of Fig.~\ref{fig:lib_amp}  shows these events, with numerous
particles cascading out of resonance together.
Figure~\ref{fig:lib_amp} also shows selected examples of some lost
Trojans (middle panel) and others that survive these events (bottom
panel).  In general, the cascading episodes that occur earlier affect
only those Trojans with relatively large libration amplitudes, with
more tightly bound particles affected by each subsequent episode.
This suggests that the cause of these events is gaining strength with
time.  Note also that during any given single cascading episode
smaller libration amplitude (higher lib.\ freq.) particles are
affected first and larger libration amplitude (lower lib.\ freq.)
particles are picked up as the episode progresses.  Also, the time
between episodes, from the beginning of one to the beginning of the
next, increases systematically by about a factor of two.  These
time-dependent features suggest that the cause has a characteristic
frequency that is slowing appreciably during the course of each
episode and from one episode to the next.  Finally, those Trojans that
survive the episodes (bottom panel, Fig.~\ref{fig:lib_amp}) experience
sharp changes in libration amplitude at times coincident with other
particles of similar libration amplitude being swept up in an event.
Figure~\ref{fig:lib_amp} is an example from a simulation with $\tau =
5 \times 10^6$ years but the general characteristics of the cascading
episodes were similar for all values of $\tau$.  The primary
difference was that the number of particles lost during each event
increased as $\tau$ increased.

Many of these features are similar to the effects of sweeping
secondary resonances studied in the tidal evolution of the uranian
satellite system (Tittermore \& Wisdom 1990; Malhotra \& Dermott 1990;
Malhotra 1998a).   In particular, the anomalously large orbital
inclination of  Miranda is naturally explained as a consequence of
secondary  resonance sweeping due to tidal evolution within a 3:1
inclination-type mean-motion resonance with Umbriel.  This  resonance
increased Miranda's inclination from an initially  small value, but
the resonance was temporary. As Miranda's  inclination approached its
current value, the satellites were  captured in a secondary resonance
which amplified their primary  resonance libration amplitude; this
eventually caused the  satellites to escape from the 3:1 mean-motion
resonance,  leaving Miranda with an inclination that is preserved to
the  present. The secondary resonance implicated in this case was  due
to a 1:3 commensurability between the libration frequency  of the
mean-motion resonance angle and the secular frequency  of precession
of the relative lines of nodes of the two  satellites.  The exact
secondary resonance can be identified  because each one is associated
with a specific value of the  final inclination of Miranda.

The essential elements of the Uranian satellite dynamics  relevant to
the present work are that (i) secondary resonances  appear at the
center of the primary resonance, initially at small  libration
amplitude, and sweep across the entire libration region of the primary
resonance; (ii) capture into a secondary resonance  is a probabilistic
phenomenon, with low-integer secondary  resonances having generally
higher probabilities of capture;  (iii) capture in a secondary
resonance causes an amplification of  the libration amplitude until
the system escapes the mean-motion  resonance; (iv) particles not
captured experience a small  perturbation, mainly in their libration
amplitude; they may later be  captured by another secondary resonance
sweeping by.  This analogy suggests that the cascading episodes of
Neptune's  Trojans found in our models (Fig.~\ref{fig:lib_amp}) are
caused by one or more secondary resonances sweeping across the phase
space occupied by the Trojan particles.  Some Trojan particles can be
trapped in these secondary resonances and forced out of Neptune's
Lagrange regions.  Others only experience a momentary perturbation as
a secondary resonance crosses their location in phase space.

During the cascading episodes shown in Fig.~\ref{fig:lib_amp} the
strongest resonance in the vicinity of Neptune and its Trojan
particles is the 1:2 mean-motion with Uranus.  While Uranus and
Neptune are not locked in 1:2 resonance with each other, they are
close.  One critical argument for the Uranus-Neptune near-resonance is
$\phi^\msct{UN}_\msct{1:2}=2\lambda_\msct{Nep} - \lambda_\msct{Ura} -
\tilde{\omega}_\msct{Ura}$, where $\tilde{\omega}_\msct{Ura}$ is the
longitude of pericenter for Uranus.  During migration
$\phi^\msct{UN}_\msct{1:2}$ circulates through all angles 0 to
$360^\circ$ while the Neptune-Trojan critical argument,
$\phi^\msct{NT}_\msct{1:1}$, librates about $\pm 60^\circ$.  The
libration frequency of $\phi^\msct{NT}_\msct{1:1}$, given as
\fnt{1:1}, is $\sim10^{-4}$ yr$^{-1}$ and decreases only slightly as
Neptune migrates outward ($\propto a_\msct{Nep}^{-3/2}$).  The
circulation frequency of $\phi^\msct{UN}_\msct{1:2}$, given as
\fun{1:2}, is initially more rapid than \fnt{1:1} but decreases
dramatically as the migrating Uranus and Neptune converge upon 1:2
resonance.  As \fun{1:2} converges toward \fnt{1:1} Trojan particles
can become trapped in secondary resonances that occur when \fun{1:2}
becomes commensurate with harmonics of \fnt{1:1}, where $\fun{1:2}
\approx (j/k) \fnt{1:1}$, $j$ and $k$ are integers, and $j \geq k$.

\mar{Fig.~\ref{fig:spectra}}

Figure~\ref{fig:spectra} shows an example from a $\tau = 10^7$ year
simulation where a Trojan particle was lost while trapped in a 1:4
commensurability between \fun{1:2} and \fnt{1:1}.  A fast Fourier
filter (FFT) was used to obtain power spectra of
$\phi^\msct{NT}_\msct{1:1}$ (middle panel) and
$\phi^\msct{UN}_\msct{1:2}$ (bottom panel).  Spectra were taken every
$0.1\,\tau$ ($10^6$ years) from 1.4 to $2.8\tau$.  Each FFT used 4096
points sampled every 100 years, giving an FFT interval of about
$0.041\tau$.  In the first interval, starting at 1.4$\tau$, \fun{1:2}
is higher than the 6th harmonic of \fnt{1:1}.  As Uranus and Neptune
converge upon the 1:2 resonance \fun{1:2} slows and passes the
$6^\msc{th}$ harmonic of \fnt{1:1} at about 1.55$\tau$.  The libration
amplitude of the Trojan particle experiences a perturbation at this
time.  Passage of \fun{1:2} by the $5^\msc{th}$ harmonic of \fnt{1:1}
at 1.85$\tau$ results in a more significant change in libration
amplitude.  At about $2.23\tau$ the $4^\msc{th}$ harmonic is reached,
where $\fun{1:2} \approx 4\fnt{1:1}$.  After $2.23\tau$, as Uranus and
Neptune migrate closer to 1:2 resonance the particle is captured in
this 1:4 secondary resonance between \fun{1:2} and \fnt{1:1}; as a
consequence it is gradually forced to larger libration amplitude.
Increasing libration amplitude results in a  slower libration
frequency.  Thus, the $4^\msc{th}$ harmonic of the slowing \fnt{1:1}
keeps pace with the slowing \fun{1:2}.  The particle is trapped and
the 1:4 secondary resonance is maintained.  Trapped particles forced
to libration amplitudes of 110 to $130^\circ$ were quickly lost from
Neptune's Lagrange regions.

Initially, \fun{1:2} is more than a factor of 10 faster than
\fnt{1:1}.  After a time of $5\tau$ Uranus and Neptune have reached
nearly their final orbits and \fun{1:2} is less than twice as fast as
\fnt{1:1} for a typical unaffected Trojan particle.  As \fun{1:2}
slows it passes numerous higher harmonics of \fnt{1:1}.  The
occurrences of commensurabilities between \fun{1:2} and harmonics of
\fnt{1:1} correspond precisely to the cascading episodes shown in
Fig.~\ref{fig:lib_amp}.  In the middle panel of Fig.~\ref{fig:lib_amp}
the last particle lost (near $4.8\tau$) was initially tightly bound to
Neptune's \L{4} or \L{5} with a libration amplitude below $10^\circ$.
This particle became trapped in a 1:2 secondary resonance, $\fun{1:2}
\approx 2\fnt{1:1}$, and was lost in about $5 \times 10^6$ years.  The
first particle lost in the examples of Fig.~\ref{fig:lib_amp} (near
$1.5\tau$) was likely trapped in the 1:11 secondary resonance.  The
1:9 and 1:10 commensurabilities are not represented in the examples
shown in Fig.~\ref{fig:lib_amp}.

\mar{Figs.~\ref{fig:lib_amp_1tau},\ref{fig:spectra_1tau}}

In all simulations significant loss of Neptune Trojans occurred in the
first multiple of $\tau$, before the first bars on the histogram of
Fig.~\ref{fig:tau_histogram}.  An example of Trojan evolution during
this early period is shown in Fig.~\ref{fig:lib_amp_1tau} for a
simulation with $\tau = 10^7$ years.  The format is similar to
Fig.~\ref{fig:lib_amp}, with a cross-section of particles shown at the
top, examples of lost Trojans in the middle panel, and two typical
survivors at the bottom.  As in Fig.~\ref{fig:lib_amp}, loss of
Trojans is confined to a  few discrete episodes, primarily from
0.12--0.24$\tau$ and 0.32--0.44$\tau$.  Because these losses occur so
early in the migration process the 3:5 and 4:7 Uranus-Neptune
mean-motion resonances are likely culprits.
Figure~\ref{fig:spectra_1tau} shows a detailed analysis of one Trojan
particle lost at $0.44\tau$.  Spectra of $\phi^\msct{NT}_\msct{1:1}$
(middle panel) are shown in comparison with spectra (bottom panel) of
a critical argument of the Uranus-Neptune 4:7 mean-motion
near-resonance, $\phi^\msct{UN}_\msct{4:7}=7\lambda_\msct{Nep} -
4\lambda_\msct{Ura} - 3\tilde{\omega}_\msct{Ura}$.  Initially, as
Uranus and Neptune approach 4:7 resonance the circulation frequency of
the critical argument, \fun{4:7}, slows and converges upon \fnt{1:1}.
Abrupt changes in the Trojan libration amplitude record passage of
\fun{4:7} by the $3^\msc{rd}$, $2^\msc{nd}$, and $1^\msc{st}$
harmonics of \fnt{1:1}.  This particular Trojan survives these events
but Fig.~\ref{fig:lib_amp_1tau} illustrates that many particles are
lost in events leading up to the 1:1 commensurability between
\fun{4:7} and \fnt{1:1}.  At about $0.27\tau$ all of the power in the
spectrum of $\phi^\msct{UN}_\msct{4:7}$ falls near zero frequency.
This indicates that circulation of $\phi^\msct{UN}_\msct{4:7}$ has
ceased, with Uranus and Neptune crossing their 4:7 resonance.  As the
planets withdraw from the 4:7 resonance circulation of
$\phi^\msct{UN}_\msct{4:7}$ resumes and speeds up, with \fun{4:7} now
diverging from \fnt{1:1}.  Divergent passages of \fun{4:7} by
harmonics of \fnt{1:1} result in abrupt changes to the particle's
libration amplitude.  The particle is finally lost following passage
of \fun{4:7} by the $8^\msc{th}$ harmonic of \fnt{1:1}.

The effects of secondary resonances demonstrated by
Figs.~\ref{fig:spectra}~and~\ref{fig:spectra_1tau} can be used to
associate gross features seen in
Figs.~\ref{fig:lib_amp}~and~\ref{fig:lib_amp_1tau} with their
respective causes.  In Fig.~\ref{fig:lib_amp_1tau} the initial loss of
Trojans between 0 and $0.06\tau$ corresponds to withdrawal of Uranus
and Neptune from the location of their 3:5 mean-motion resonance.
Between 0.12 and $0.24\tau$ the planets are approaching their 4:7
resonance.  The lull from 0.24 to $0.3\tau$ corresponds to a calm
period when the planets actually cross the location of their 4:7
resonance.  Withdrawal from the 4:7 leads to more lost Trojan
particles between 0.3 and $0.44\tau$.  Structure from 0.44 to
$0.58\tau$ is from approach to the 5:9, with withdrawal from 5:9
between 0.6 and $0.8\tau$.  We can also see structure associated with
approach to and withdrawal from the 6:11 resonance, between 0.8 and
$1.1\tau$ (more evident in Fig.~\ref{fig:lib_amp}).  The two survivors
shown as examples in the bottom panel of Fig.~\ref{fig:lib_amp_1tau}
show sudden changes in libration amplitude probably caused by
secondary resonances associated with the 6:11 resonance.

Loss of Neptune Trojans during secondary resonances with Uranus and
Neptune occurred for all values of $\tau$.  However, the fraction lost
during passages of each secondary resonance increased as $\tau$
increased.  Figure~\ref{fig:tau_histogram} essentially gives the
results for loss during approach to the Uranus-Neptune 1:2 mean-motion
resonance.  With $\tau = 10^6$ years about 50\% of the Trojan
particles still present are lost through secondary resonances
associated with the 1:2.  With $\tau = 10^7$ years about 85\% of
existing Trojans are lost during these same events.  For times less
than $\tau$ Fig.~\ref{fig:tau_histogram} provides bulk survival
statistics for events associated with higher order Uranus-Neptune
resonances such as the 3:5, 4:7, and 5:9.  These include withdrawal
from the 3:5 as well as approach and withdrawal from both the 4:7 and
5:9.  With $\tau = 10^6$ years about 20\% of Trojan particles are lost
through secondary resonances associated with all these events.  In
contrast, with $\tau = 10^7$ years nearly 65\% of Trojan particles are
lost during the same events.

Most of the Trojans lost in our simulations were destabilized by
perturbations from Uranus.  The primary episodes of loss can be
directly linked to the secondary resonances described above.  The
nearest saturnian and jovian mean-motion resonances to Neptune are the
saturnian 1:6 and jovian 1:13 and 1:14.  Direct perturbations of
Jupiter and Saturn on Neptune Trojans appear to play a very minor
role, as indicated by results from the simulation run without Uranus.
Indirect perturbations of Jupiter and Saturn, acting through Uranus,
may have a greater influence.  One important mean-motion resonance in
the region of Uranus is the jovian 1:7.  The  pair cross their 1:7
mean-motion resonance between 3.25$\tau$ and 3.5$\tau$ (see
Fig.~\ref{fig:planets}).  Figure~\ref{fig:lib_amp} shows that some
Trojan particles trapped in the 1:3 secondary resonance ($\fun{1:2}
\approx 3\fnt{1:1}$) are lost in this time interval.  Spectral
analysis of some of these trapped particles indicates that they can
escape the secondary resonance during passage of the Jupiter-Uranus
1:7 mean-motion resonance.  It is possible then that some of the
Trojan particles forced to high libration amplitudes while trapped in
the 1:3 secondary resonance could ultimately have been lost as a
result of the Jupiter-Uranus 1:7 mean-motion resonance.

\vspace{-0.5cm}
\section*{\centering {\normalsize III. DISCUSSION}}
\vspace{-0.2cm}

\mar{Figs.~\ref{fig:final_ele},\ref{fig:ecc_tau}}

The post-formation orbital evolution of the giant planets in our solar
system is constrained by observed characteristics of small body
populations.  Studies of the interaction of the giant planets with
these populations have led to a detailed understanding of the early
dynamical evolution of the outer solar system.  Discovery of the first
Trojan companion of Neptune, 2001 QR\,322, reveals yet another class
of minor solar system body.  As more of these objects are discovered,
statistical comparisons of their orbits with results from numerical
modeling such as ours will provide further constraints on the early
history of the solar system.  With only a single known Neptune Trojan,
such a comparison may not carry much weight but is still interesting.
Figure~\ref{fig:final_ele} shows the initial and final distributions
of eccentricity, inclination, and libration amplitude for surviving
Trojans from the $\tau = 10^7$ year simulation.  The initial and final
distributions of inclination and libration amplitude show no strong
differences.  The eccentricity distributions are markedly different
though.  Figure~\ref{fig:ecc_tau} shows the mean eccentricity of
surviving Trojan particles as a function of time for all values of
$\tau$.  The initial eccentricities of the particles (at {\sf
T}$_\msc{o} = 25,000$ years) were about equal to those of Jupiter's
real Trojan asteroids, with a mean of 0.075 and standard deviation of
about $\pm0.03$.  The secondary resonances primarily removed higher
eccentricity particles and longer migration time scales led to erosion
of lower eccentricity particles.  With $\tau = 10^7$ years the
surviving population at $5\tau$ has a mean eccentricity of
$0.04\pm0.02$.

The ($a,e,i$) space occupied by the surviving Neptune Trojans in our
$\tau = 10^7$ years simulation overlaps remarkably well with the large
stable region found by Nesvorn\'y and Dones (2002).  In their study,
using the present configuration of the planets as starting conditions,
nearly all Trojan particles within $\pm 0.1$ AU of the libration
center and with $e \leq 0.07$, $i \leq 25^\circ$, $A \leq 60-70^\circ$
survived for $4 \times 10^9$ years.  The agreement is probably not
coincidental considering the importance we found for the 1:2
mean-motion near-resonance between Uranus and Neptune.  Nesvorn\'y and
Dones found that this same near-resonance shapes the stability of
Neptune's Trojan regions in the present planetary configuration.  The
similarity between our final conditions for $\tau = 10^7$ years and
the initial conditions of Nesvorn\'y and Dones suggests that 30 to
50\% of Neptune's post-migration Trojan population could survive for
the age of the solar system.  This indicates that Neptune's current
Trojan population may be just 1-2\% of its primordial size.

Depending on initial conditions it may be possible to either mitigate
or enhance Trojan losses during planetary migration.  From the
characteristics of Pluto's orbit, Malhotra (1995) shows that 5\,AU
should be considered a minimum migration distance for Neptune.  This
smaller distance could allow Neptune Trojans to avoid destabilizing
secondary resonances associated with withdrawal from the 3:5 with
Uranus and some of those associated with approach to the 4:7.  On the
other hand, Gomes (1997) suggests that Neptune may have migrated as
much as 9--10\,AU rather than the nominal value of 7\,AU assumed in
our simulations.  The extra 2\,AU would put Neptune initially interior
to the location of the 3:5 mean-motion resonance with Uranus.  Neptune
Trojans would then be subject to additional losses through secondary
resonances associated with approach to the 3:5.  Convergence of Uranus
and Neptune toward 1:2 mean-motion resonance seems unavoidable based
on our current understanding of planetary migration.  Therefore,
regardless of the extent of Neptune's migration (be it 5, 7, or
9\,AU), it is the migration rate on final approach to its present
orbit that determined the fate of most of its Trojans.

Previous work (Malhotra 1998b) suggests that planetary migration with
$\tau = 10^7$ years is favored to allow Kuiper belt objects trapped in
2:3 mean-motion resonance with Neptune to evolve to the high orbital
inclinations observed for some of the plutinos.  If Neptune had an
abundant primordial population of Trojan companions then migration
with $\tau \geq 10^7$ years would have dramatically depleted that
population.  The loss rate would not have been a smooth function of
time.  Instead, the losses would have been episodic, with an
especially dramatic loss between 4 and 5$\tau$ as Uranus and Neptune
closed upon their present near-resonant configuration.  Our
simulations show that this final loss episode involved a nearly 75\%
reduction in Neptune's Trojan population (see
Fig.~\ref{fig:tau_histogram}).  These liberated bodies, heretofore
dynamically sequestered by Neptune, would be injected into the largely
eroded planetesimal disk.  We speculate that if their numbers were
sufficiently high they could have temporarily enhanced the impact rate
throughout the solar system.  Traces of this event, while probably not
preserved in Triton's relatively fresh surface, may be preserved in
the impact records on other more ancient surfaces.

\vspace{-0.5cm}
\section*{\centering {\normalsize IV. ACKNOWLEDGMENTS}}
\vspace{-0.2cm}

We thank Bill Hartmann and Stu Weidenschilling for valuable comments
on an early draft of this paper.  This material is based upon work
supported by the National Aeronautics and Space Administration under
Grant Nos.\ NAG5-10343 and NAG5-11661 issued through the Office of
Space Science and by the Research Foundation of S\~ao Paulo State
(FAPESP) under Grant No.\ 01/13447-6.

\vspace{-0.5cm}
\section*{\centering { \normalsize V. REFERENCES}}
\vspace{-0.2cm}


\newcommand{\vpy}[3]{\textbf{#1}, #2--#3.}

\newcommand{\science}[4]{\textsl{Science} \vpy{#1}{#2}{#3}}
\newcommand{\icarus}[4]{\textsl{Icarus} \vpy{#1}{#2}{#3}}
\newcommand{\icarusip}[4]{\textsl{Icarus} (in press).}
\newcommand{\icarussub}[4]{\textsl{Icarus} (submitted).}
\newcommand{\nature}[4]{\textsl{Nature} \vpy{#1}{#2}{#3}}
\renewcommand{\apj}[4]{\textsl{Astrophys.\ J.} \vpy{#1}{#2}{#3}}
\newcommand{\apjip}[1]{\textsl{Astrophys.\ J.} (in press).}
\newcommand{\apjsub}[4]{\textsl{Astrophys.\ J.} (submitted).}
\renewcommand{\apjl}[4]{\textsl{Astrophys.\ J. Lett.\ }\vpy{#1}{#2}{#3}}
\newcommand{\apjlip}[1]{\textsl{Astrophys.\ J. Lett.\ }(in press).}
\newcommand{\apjlsub}[4]{\textsl{Astrophys.\ J. Lett.\ }(submitted).}
\newcommand{\ana}[4]{\textsl{Astron.\ Astrophys.\ }\vpy{#1}{#2}{#3}}
\renewcommand{\aj}[4]{\textsl{Astron.\ J.} \vpy{#1}{#2}{#3}}
\newcommand{\ajip}[1]{\textsl{Astron.\ J.} (in press).}
\renewcommand{\araa}[4]{\textsl{Annu.\ Rev.\ Astron.\ Astrophys.\ }\vpy{#1}{#2}{#3}}     
\newcommand{\astross}[4]{\textsl{Astrophys.\
Space Sci.\ }\vpy{#1}{#2}{#3}}     
\newcommand{\iauc}[2]{\textsl{IAU Circulars} #1.}     
\renewcommand{\pasp}[4]{\textsl{Pub.\ Astron.\ Soc.\ Pacific} \vpy{#1}{#2}{#3}}  
\newcommand{\pnas}[4]{\textsl{Proc.\
Natl.\ Acad.\ Sci.\ }\vpy{#1}{#2}{#3}}
\renewcommand{\mnras}[4]{\textsl{Mon.\ Not.\ Roy.\ Astron.\ Soc.\ }\vpy{#1}{#2}{#3}}  
\newcommand{\moonpl}[4]{\textsl{Moon
Planets}\vpy{#1}{#2}{#3}}   \newcommand{\mpc}[2]{\textsl{Minor Planet
Circulars} #1.}    \newcommand{\mpec}[2]{\textsl{Minor Planet Ele.\
Circulars} #1.}   \newcommand{\ptp}[4]{\textsl{Prog.\ Theoret.\ Phys.\
}\vpy{#1}{#2}{#3}}    \newcommand{\pia}[4]{\textsl{Proc.\ Irish Acad.\
}\vpy{#1}{#2}{#3}}    \newcommand{\pss}[4]{\textsl{Plan. Spa. Sci.\
}\vpy{#1}{#2}{#3}}    \newcommand{\cmda}[4]{\textsl{Celes.\ Mech.\
Dynamic.\ Astron.\ }\vpy{#1}{#2}{#3}}
\newcommand{\an}[4]{\textsl{Astronomische
Nachrichten}\vpy{#1}{#2}{#3}} \newcommand{\lpsc}[2]{\textsl{Lunar
Planet.\ Sci.\ Conf.\ } {\bf #1}, abstract \# #2 on CD-ROM.}
\renewcommand{\phd}[2]{\textsl{Ph.D. dissertation}, #1, #2.}

\newcommand{\astii}[2]{In \textsl{Asteroids II},  (R.P. Binzel,
			T. Gehrels, and M.S. Matthews, Eds.),
			pp.~#1--#2. Univ.\ Arizona Press, Tucson.}

\newcommand{\ppiii}[2]{In \textsl{Protostars and Planets III},
			(E.H.~Levy and J.I.~Lunine, Eds.),
			pp.~#1--#2. Univ.\ Arizona Press, Tucson.}

\newcommand{\ppiv}[2]{In \textsl{Protostars and Planets IV},
			(V.~Mannings, A.P.~Boss, and S.S.~Russell,
			Eds.)  pp.~#1--#2. Univ.\ Arizona Press,
			Tucson.}

\newcommand{\oem}[2]{In \textsl{Origin of the Earth and Moon},
			(R.M.~Canup and K.~Righter, Eds.),
			pp.~#1--#2. Univ.\ Arizona Press, Tucson.}

\newcommand{\meoc}[2]{In \textsl{Motion, Evolution of Orbits, and
			Origin of  Comets. Proceedings, IAU Symposium
			No. 45 Leningrad},  (E.I. Chevotarev et al.,
			Eds.), pp.~#1--#2. Springer-Verlag, New
			York/Berlin.}

\newcommand{\neptune}[2]{In \textsl{Neptune and Triton},
			(D.P. Cruikshank, Ed.), pp.~#1--#2. Univ.\
			Arizona Press, Tucson.}

\newcommand{\pluto}[2]{In \textsl{Pluto and Charon},  (S.A. Stern and
			D.J. Tholen, Eds.), pp.~#1--#2. Univ.\ Arizona
			Press, Tucson.}

\newcommand{\soj}[2]{In \textsl{Satellites of Jupiter},  (D. Morrison,
			Ed.), pp.~#1--#2. Univ.\ Arizona Press,
			Tucson.}

\newcommand{\iss}[2]{In \textsl{Ices in the Solar System},
			(J. Klinger, D. Benest, A. Dollfus,
			R.D. Smoluchowski, Eds.), pp.~#1--#2. Reidel,
			Dordrecht.}

\begin{list}{}{ \setlength{\leftmargin}{0.5cm}
		\setlength{\itemindent}{-0.5cm}}

\item {\sc Chen, J., D. Jewitt, C. Trujillo, and J. Luu} 1997.  Mauna
Kea Trojan survey and statistical studies of \L{4} Trojans.  {\it
Bull.\ Amer.\ Astron.\ Soc.\ }{\bf 29}, DPS abstract \#25.08.
  
\item {\sc Chiang, E.I., and A.B. Jordan} 2002.  On the plutinos and
twotinos of the Kuiper belt.  \aj{124}{3430}{3444}{2002}
  
\item {\sc Chiang, E.I.} 2003.   2001 QR\,322.   \iauc{8044}
  
\item {\sc Cordeiro, R.R., R.S. Gomes, and R.V. Martins} 1997.  A
mapping for nonconservative systems.  \cmda{65}{407}{419}{1997}

\item {\sc de la Barre, C.M., W.M. Kaula, and F. Varadi} 1996.  A
study of orbits near Saturn's triangular Lagrangian points.
\icarus{121}{88}{113}{1996}

\item {\sc Fernandez, J.A., and W.H. Ip} 1984.   Some dynamical
aspects of the accretion of Uranus and Neptune:   The exchange of
orbital angular momentum with planetesimals.
\icarus{58}{109}{120}{1984}
  
\item {\sc Fernandez, J.A., and W.H. Ip} 1996.   Orbital expansion and
resonant trapping during the   late accretion stages of the outer
planets.   \pss{44}{431}{439}{1996}

\item {\sc Fleming, H.J., and D.P. Hamilton} 2000.   On the origin of
the Trojan asteroids:   Effects of Jupiter's mass accretion and radial
migration.   \icarus{148}{479}{493}{2000}

\item {\sc Friedland, L.} 2001.  Migration timescale thresholds for
resonant  capture in the plutino problem.  \apjl{547}{75}{79}{2001}

\item {\sc Gomes, R.S.,} 1997.  Orbital evolution in resonance lock I.
The restricted 3-body problem.  \aj{114}{2166}{2176}{1997}

\item {\sc Gomes, R.S.,} 1998.   Dynamical effects of planetary
migration on  primordial Trojan-type asteroids.
\aj{116}{2590}{2597}{1998}

\item {\sc Gomes, R.S.,} 2000.  Planetary migration and plutino
orbital inclinations.  \aj{120}{2695}{2707}{2000}

\item {\sc Gomes, R.S.} 2003.  
	The origin of the Kuiper belt high-inclination population.
	\icarus{161}{404}{418}{2003}

\item {\sc Hahn, J.M., and R. Malhotra}  1999.    Orbital evolution of
planets embedded in a planetesimal disk.   \aj{117}{3041}{3053}{1999}

\item {\sc Holman, M.J., and J. Wisdom} 1993.  
 	Dynamical stability in the outer solar system and the  
 	delivery of short period comets.
 	\aj{105}{1987}{1999}{1993}

\item {\sc Liou, J.C., and R. Malhotra} 1997.   
	Depletion of the outer asteroid belt.   
	\science{275}{375}{377}{1997}

\item {\sc Malhotra, R.,} 1993.   The origin of Pluto's peculiar
orbit.  \nature{365}{819}{821}{1993}
 
\item {\sc Malhotra, R.,} 1995.   The origin of Pluto's orbit:
Implications for the solar system beyond Neptune.
\aj{110}{420}{429}{1995}
 

\item {\sc Malhotra, R.,} 1998a. 
  Orbital resonances and chaos in the solar system,
  In {\it Solar System Formation and Evolution},
  (D. Lazzaro, R. Vieira Martins, S. Ferraz-Mello, J. Fern\'andez, 
  and C. Beaug\'e, Eds.), ASP Conference Series {\bf 149},
  pp.~37--63.
 
\item {\sc Malhotra, R.,} 1998b.   
	Pluto's inclination excitation by resonance sweeping.
	\lpsc{29}{1476}
 
\item {\sc Malhotra, R., and S.F. Dermott} 1990.    The role of
secondary resonances in the orbital history of Miranda.
\icarus{85}{444}{480}{1990}
 
\item {\sc Malhotra, R., M. Duncan, and H. Levison} 2000.   Dynamics
of the Kuiper belt.   \ppiv{1231}{1254}

\item {\sc Marsden, B.G.} 2003.   2001 QR\,322.  \mpec{2003-A55}

\item {\sc Michtchenko, T.A., C. Beaug\'e, and F. Roig} 2001.
Planetary migration and the effects of mean motion resonances  on
Jupiter's Trojan asteroids.  \aj{122}{3485}{3491}{2001}

\item {\sc Nesvorn\'y, D., and L. Dones} 2002.   How long-lived are
the hypothetical Trojan  populations of Saturn, Uranus, and Neptune?
\icarus{160}{271}{288}{2002}


\item {\sc Tittemore, W.C., and J. Wisdom} 1990.
	Tidal evolution of the uranian satellites III.
	Evolution through the Miranda-Umbriel 3:1, Miranda-Ariel 5:3,
        and Ariel-Umbriel 2:1 mean-motion commensurabilities.
	\icarus{85}{394}{443}{1990}

\item {\sc Weissman, P.R., and H.F. Levison} 1997.   The population of
the trans-neptunian region: The Pluto-Charon environment.
\pluto{559}{604}

\item {\sc Wisdom, J., and M. Holman} 1991.  Symplectic maps for the
N-body problem.  \aj{102}{1528}{1538}{1991}

\item {\sc Wisdom, J., and M. Holman} 1992.  Symplectic maps for the
N-body problem: Stability analysis.  \aj{104}{2022}{2029}{1992}
 
\mylabel{last_ms_page}

\end{list}

\newpage
  
\newcounter{tabs}

\input{table1}
\clearpage






	\gdef\capplanets{Four panels showing examples of the evolution
	with time of the semi-major axes of the four giant planets in
	a migration simulation.  The planets were subject to mutual
	gravitational perturbations and a drag force which caused
	their orbits to migrate---Jupiter inward; Saturn, Uranus, and
	Neptune outward.  Table~\ref{tab:model_parameters} gives the
	initial orbital parameters for the planets.  Time is expressed
	in units of $\tau$, the characteristic migration time scale.
	After a time of $5\tau$ migration is 99.33\% complete and
	subsequent orbital evolution is dominated by mutual planetary
	gravitational perturbations rather than the migration force.
	Dashed lines indicate the ideal locations (for zero
	eccentricity) of the 3:5 and 1:2 mean-motion resonances with
	Uranus and the 1:7 mean-motion resonance with Jupiter.}



	\gdef\captausix{Survival of Neptune Trojans in a simulation
	with the four giant planets migrating with a characteristic
	time scale of $\tau = 10^6$ years.  The six panels show
	minimum and maximum limits on libration of the critical
	argument, $\phi^\msct{NT}_\msct{1:1} = \lambda -
	\lambda_\msct{Nep}$.  The dashed lines at $\pm 60^{\circ}$
	indicate the ideal locations of the leading \L{4} and trailing
	\L{5} Lagrange equilibrium points (for zero eccentricity).
	Each histogram gives the number of surviving Trojans in each
	bin as a fraction of the total population across all bins at
	the initial time {\sf T}$_\msc{o} = 25,000$ years.  The top
	panel shows the initial distribution at time {\sf
	T}$_\msc{o}$.  Each of the subsequent lower panels lists the
	output time in units of $\tau$ and the number of particles
	trapped in the \L{4} and \L{5} regions.}



\begin{figure}  \refstepcounter{figs} \label{fig:tau_6}
	\vspace{-1cm} 
	\centerline{ 
	\hspace{-1cm}  
	\epsfxsize=6.5in  
	\epsffile[18 144 592 718]{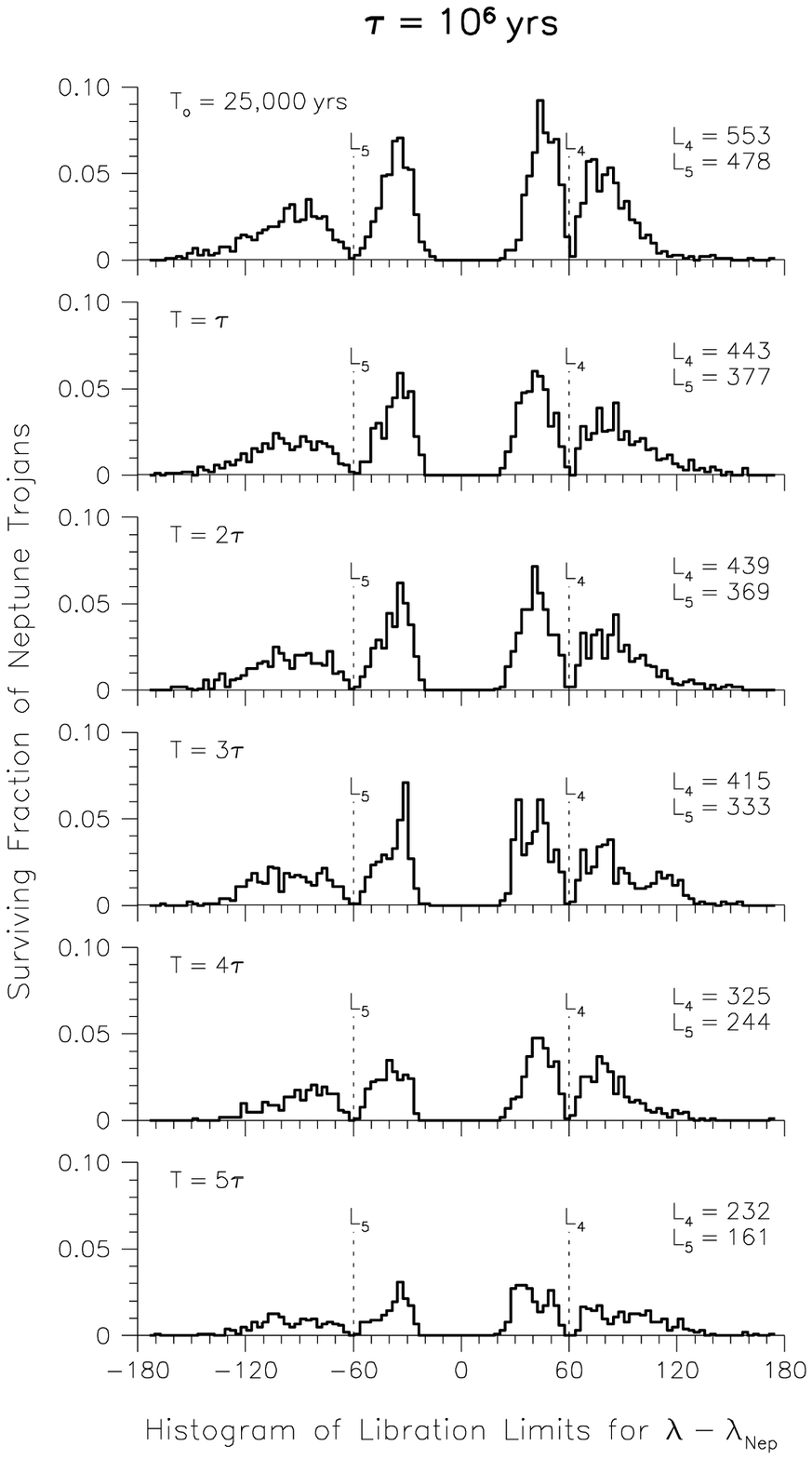}}
	\vspace{1.5cm}
	Figure \ref{fig:tau_6}: 
	\captausix
\end{figure}

	\gdef\captauseven{Similar to Fig.~\ref{fig:tau_6} but from a
	simulation with characteristic time scale 10 times slower,
	$\tau = 10^7$ years.  All other initial conditions were
	identical to the simulations shown in Fig.~\ref{fig:tau_6}.
	In this slow migration simulation nearly 99\% of Neptune's
	initial Trojan companions are lost after $5\tau$.}



	\gdef\captauhist{Histogram of the surviving fraction of
	Neptune Trojans as a function of time for four different
	characteristic migration time scales, $\tau$, ranging from
	$10^6$ to $10^7$ years.  In each case, time is expressed as
	integer multiples of $\tau$ up to $5\tau$.  Solid and open
	bars indicate surviving fractions of Trojans in Neptune's
	leading \L{4} and trailing \L{5} regions, respectively.  For
	each value of $\tau$, simulations with identical initial
	conditions were run on four different types of processors, two
	Intel Xeons (1.7 and 2.4 GHz), a Digital Alpha\,EV6, and a Sun
	Sparc Ultra\,5.  The results shown in the histogram are a
	composite from all four simulations.}


 
 	\gdef\capjsn{Similar to Fig.~\ref{fig:tau_6} ($\tau = 10^7$)
	but from a simulation that included Jupiter, Saturn, and
	Neptune, but not Uranus.  Removing Uranus resulted in nearly
	full retention of Neptune Trojans up to a time of $5\tau$.}
 


	\gdef\caplibamp{The libration amplitude of Neptune Trojans as
	a function of time expressed in units of $\tau$ for a
	simulation with $\tau = 5 \times 10^6$ years.  Evolution of
	the general population is shown in the top panel.  To avoid
	saturation only every fifth particle is plotted and libration
	amplitudes greater than 180$^\circ$ are not connected.  After
	a time of $\tau$, loss of Trojans is generally marked by
	cascading of particles out of resonance during a discrete
	number of prolonged events.  Examples of Trojans lost during
	these episodes are shown in the middle panel.  Trojans not
	lost by this process are still perturbed and can have their
	libration amplitudes decreased, as seen in the descending
	step-like appearance of the libration amplitudes of the
	survivors shown in the bottom panel.}



	\gdef\capspectra{Evolution of the libration amplitude for one
	of the lost Trojans from a $\tau = 10^7$ year simulation is
	shown here at high resolution (top panel).  A fast Fourier
	transform (FFT) was used to obtain power spectra of critical
	arguments for both the Neptune-Trojan 1:1 resonance
	($\phi^\msct{NT}_\msct{1:1}=\lambda-\lambda_\msct{Nep}$,
	middle panel) and the Uranus-Neptune 1:2 near-resonance
	($\phi^\msct{UN}_\msct{1:2}=2\lambda_\msct{Nep} -
	\lambda_\msct{Ura} - \tilde{\omega}_\msct{Ura}$, bottom
	panel).  Each FFT used 4096 points sampled every 100 years
	(see FFT interval bar), resulting in a Nyquist critical
	frequency (5 kyr$^{-1}$) well removed from the range shown
	here.  The start times of the FFTs are indicated to the right
	of the spectra.  Spectra of $\phi^\msct{NT}_\msct{1:1}$
	(middle panel) are shown in units of log power in order to
	simultaneously resolve the fundamental frequency (\fnt{1:1})
	and its higher harmonics.  The initial positions of the first
	six harmonics of \fnt{1:1} are indicated by the dashed lines.
	As \fun{1:2} converges toward \fnt{1:1} it overtakes the
	$6^\msc{th}$ and then the $5^\msc{th}$ harmonics of \fnt{1:1}.
	The Trojan particle experiences sudden changes in libration
	amplitude during these passages (indicated in top panel).  At
	about $2.23\,\tau$ (see spectra taken at $2.2\,\tau$) the
	$4^\msc{th}$ harmonic of \fnt{1:1} becomes locked to
	\fun{1:2}.}



\begin{figure}  \refstepcounter{figs} \label{fig:lib_amp}
	\vspace{-2cm} 
	\centerline{ 
	\hspace{-1cm}  
	\epsfxsize=6.5in  
	\epsffile[18 144 592 718]{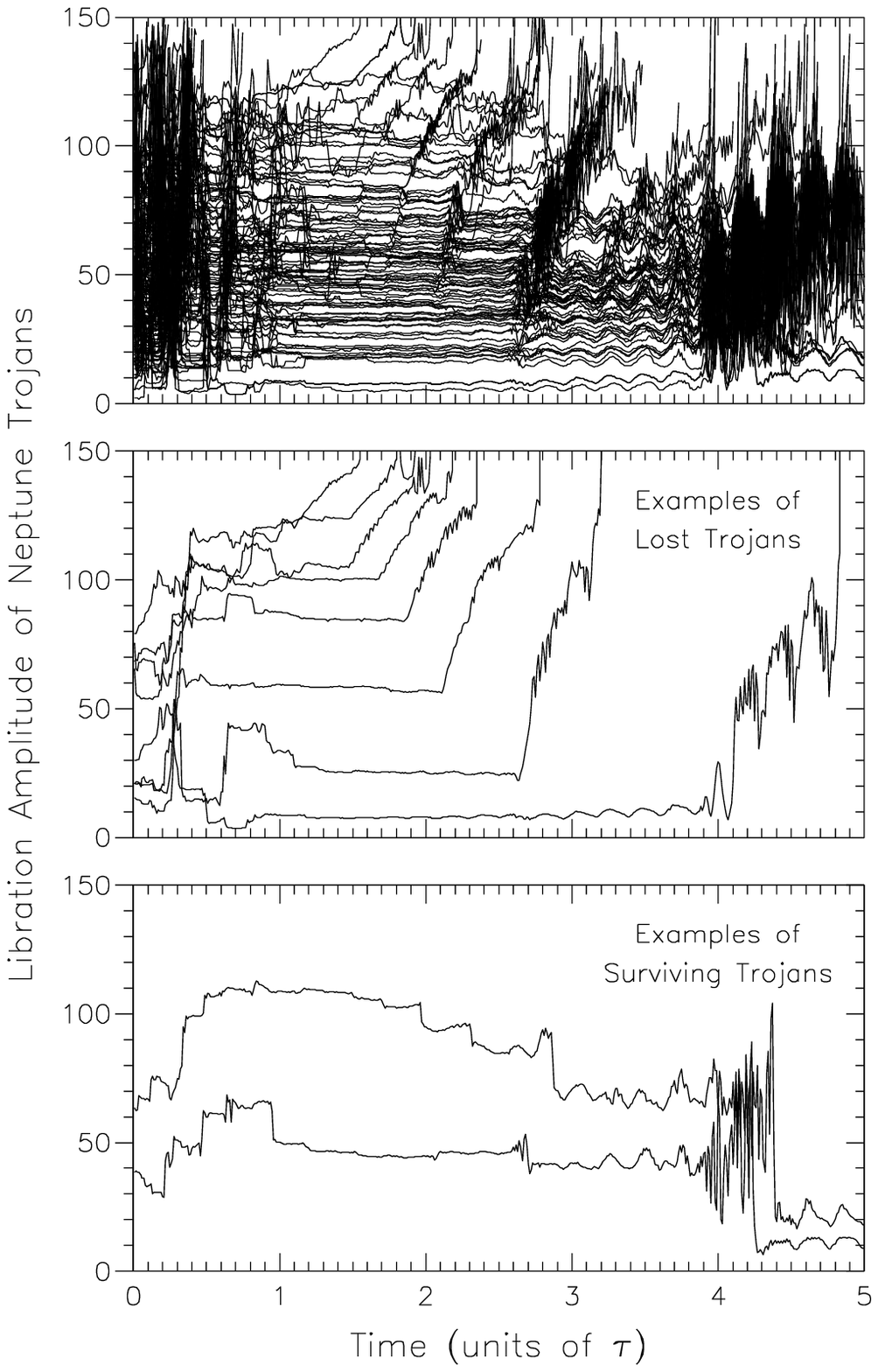}}
	\vfill
	Figure \ref{fig:lib_amp}: 
	\caplibamp
\end{figure}

	\gdef\caplibamponetau{Similar to Fig.~\ref{fig:lib_amp} except
	shown at  higher temporal resolution for the period from time
	0 to $\tau$.}



\begin{figure}  \refstepcounter{figs} \label{fig:spectra}
	\vspace{-0.5cm} 
	\centerline{ 
	\hspace{-1cm}  
	\epsfxsize=6.5in  
	\epsffile[18 144 592 718]{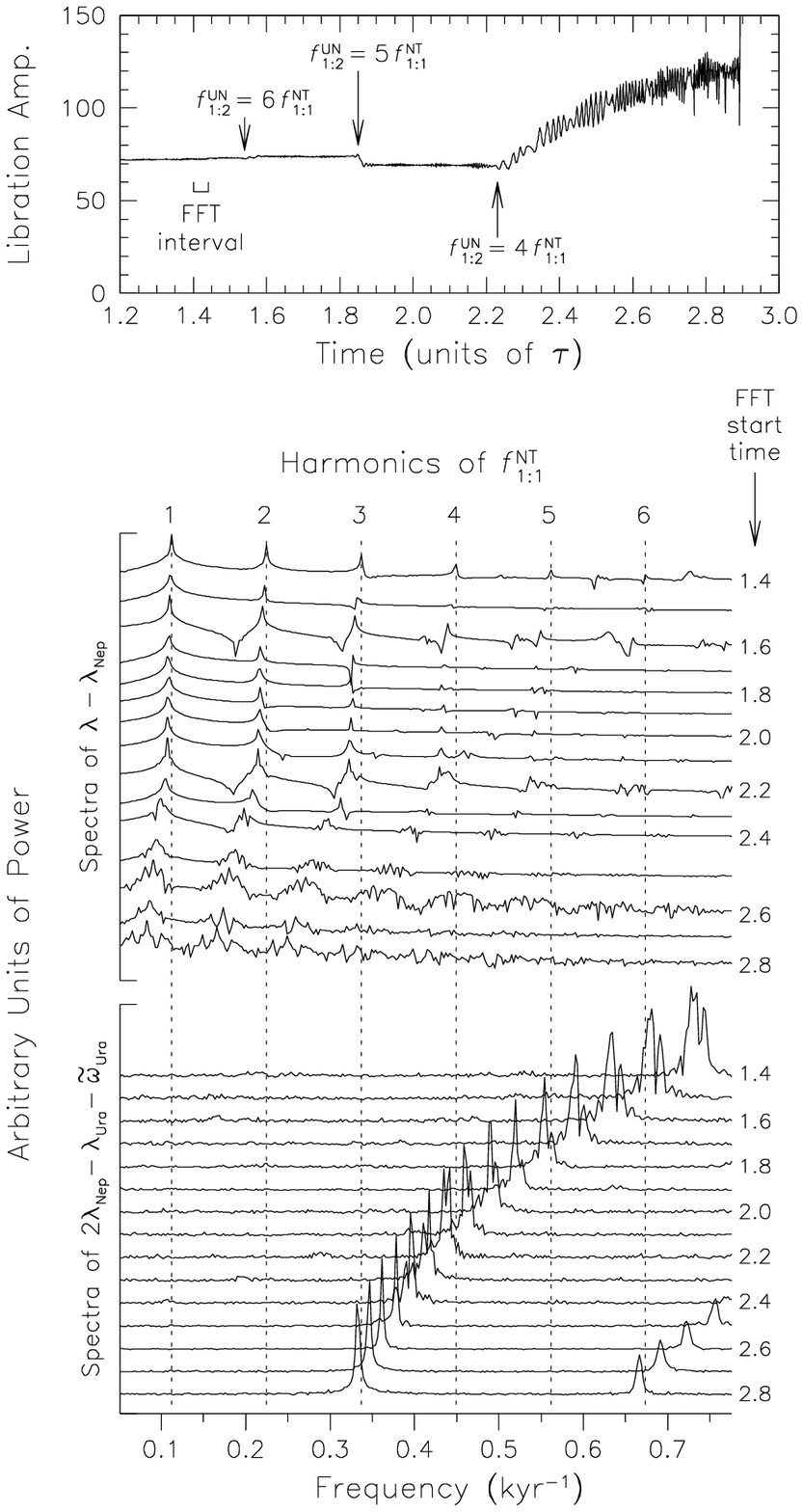}}
	\vspace{0.5cm}
	Figure \ref{fig:spectra}: 
	\capspectra
\end{figure}

\begin{figure}  \refstepcounter{figs} \label{fig:lib_amp_1tau}
	\vspace{-2cm} 
	\centerline{ 
	\hspace{-1cm}  
	\epsfxsize=6.5in  
	\epsffile[18 144 592 718]{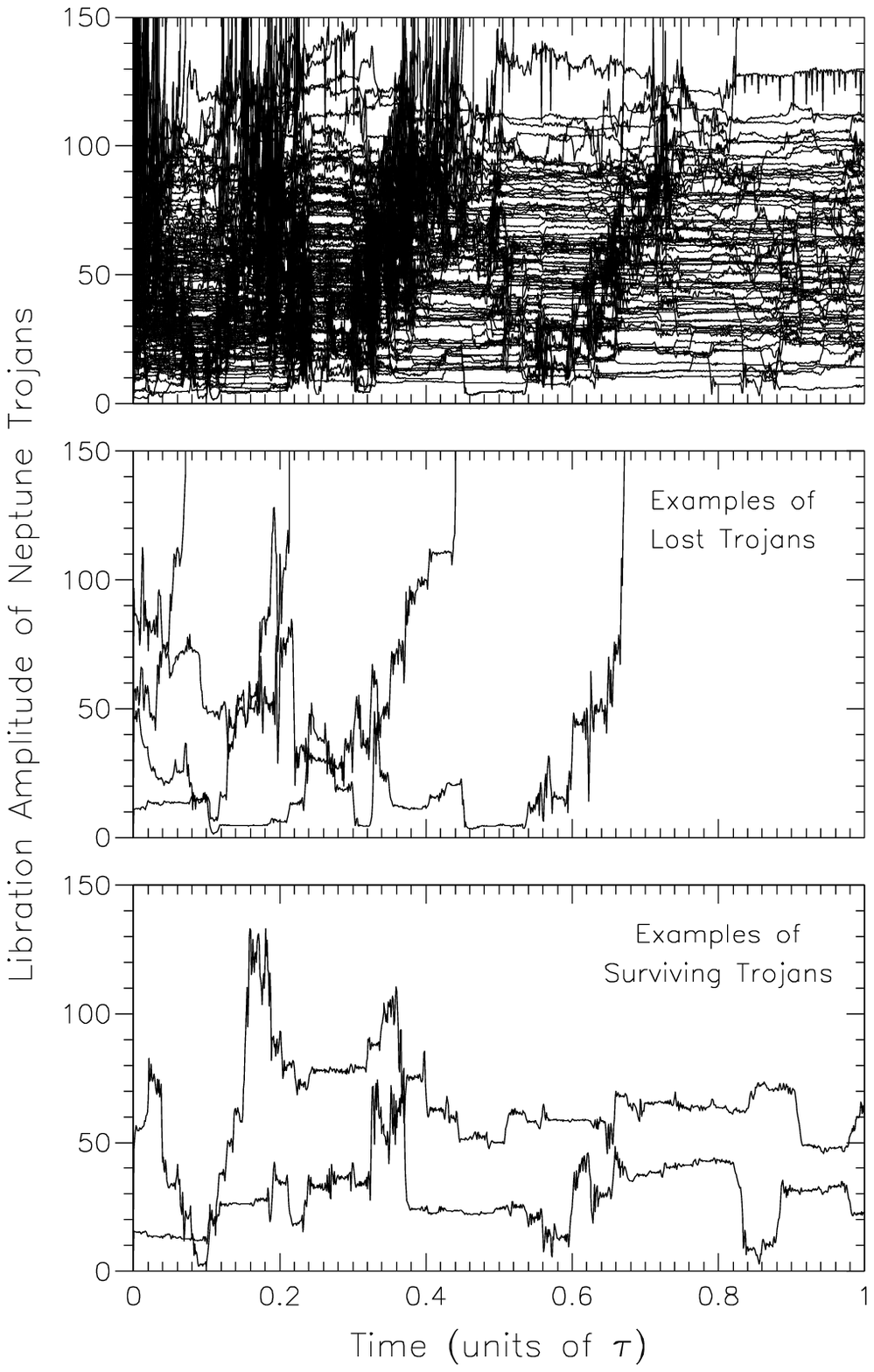}}
	\vfill
	Figure \ref{fig:lib_amp_1tau}: 
	\caplibamponetau
\end{figure}

	\gdef\capspectraonetau{Similar to Fig.~\ref{fig:spectra}.  The
	libration amplitude of one Trojan particle from
	Fig.~\ref{fig:lib_amp_1tau} is shown at higher resolution (top
	panel).  Power spectra of $\phi^\msct{NT}_\msct{1:1}$ (middle
	panel, in log power) are compared with spectra of a critical
	argument for the 4:7 mean-motion resonance with Uranus
	($\phi^\msct{UN}_\msct{4:7}=7\lambda_\msct{Nep} -
	4\lambda_\msct{Ura} - 3\tilde{\omega}_\msct{Ura}$).  Spectra
	were taken every $0.02\tau$ ($2\times10^5$ years) from 0.21 to
	$0.43\tau$ (labeled to right of spectra).  Each FFT used 1024
	points sampled every 100 years, giving an FFT interval
	(indicated in top panel) of about $0.01\tau$.  Note that
	dashed lines only indicate initial locations of the first six
	harmonics of \fnt{1:1}.  By the later spectra, \fnt{1:1} has
	slowed sufficiently to move the $8^\msc{th}$ harmonic into the
	frame.}

				


	\gdef\capfinalele{Histograms showing the distributions of
	eccentricity, inclination, and libration amplitude from the
	$\tau = 10^7$ year simulations.  The dashed line indicates the
	initial distribution at {\sf T}$_\msc{o} = 25,000$ years and
	the solid line is the final distribution after $5\tau$.  {\sf
	N} indicates the number of Trojan particles remaining after
	{\sf T}$_\msc{o}$ and $5\tau$.  The orbital eccentricity of
	2001 QR\,322 is also indicated.}



        \gdef\capecctau{Distribution of orbital eccentricity of the
        surviving  Neptune Trojans as a function of time for four
        different characteristic migration time scales, $\tau$,
        ranging from $10^6$ to $10^7$ years.  Points are the mean and
        error bars are one standard deviation.  The orbital
        eccentricity of 2001 QR\,322 is shown by the dashed line.}



\newcounter{figs}

\newpage
 
\begin{figure}  \refstepcounter{figs} \label{fig:planets}
	\vspace{-1cm} 
	\centerline{ 
	\hspace{-1cm}  
	\epsfxsize=6.5in  
	\epsffile[18 144 592 718]{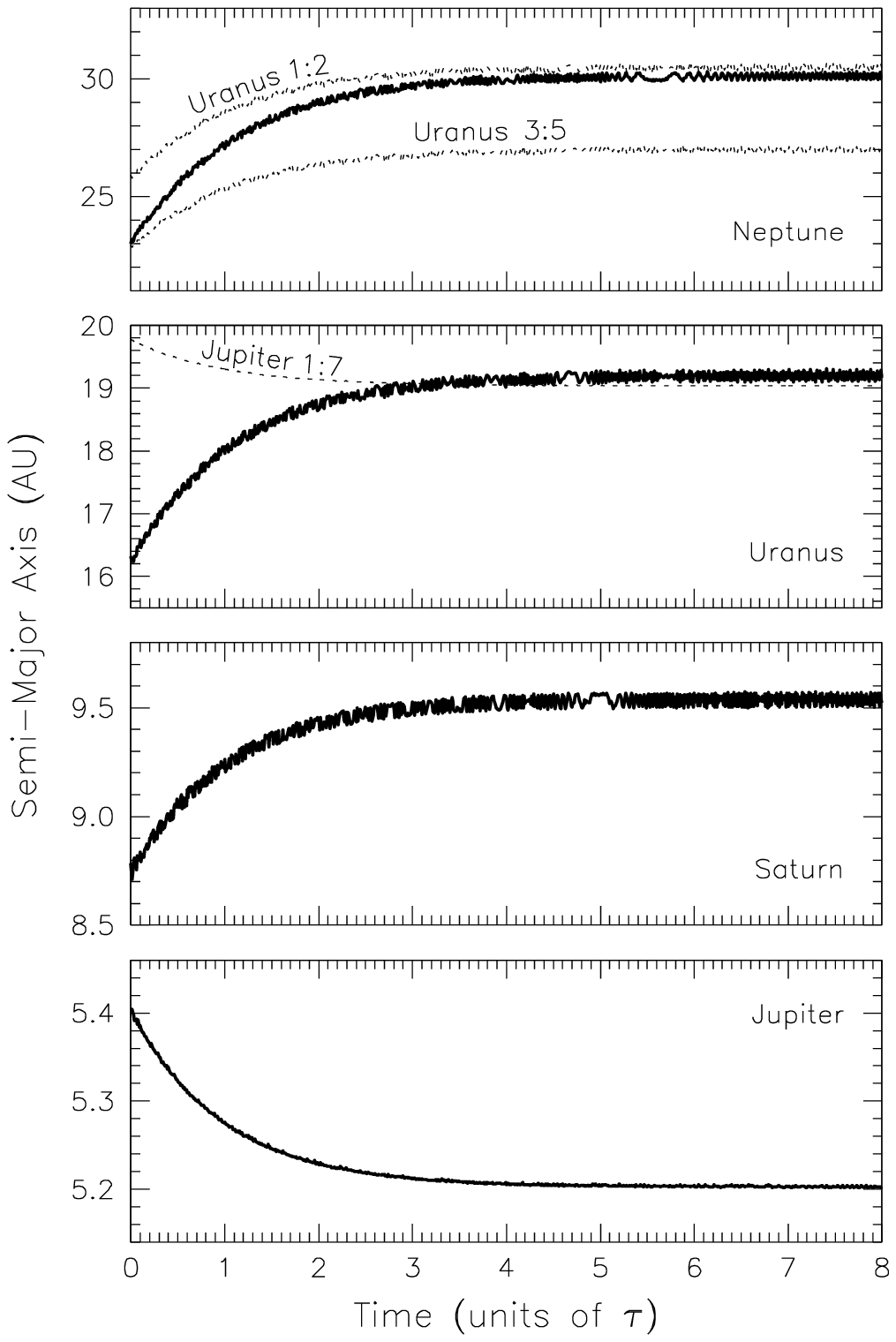}}
	\vfill
	Figure \ref{fig:planets}: 
	\capplanets
\end{figure}
\clearpage
 
\clearpage
 
\begin{figure}  \refstepcounter{figs} \label{fig:tau_7}
	\vspace{-2cm} 
	\centerline{ 
	\hspace{-1cm}  
	\epsfxsize=6.5in  
	\epsffile[18 144 592 718]{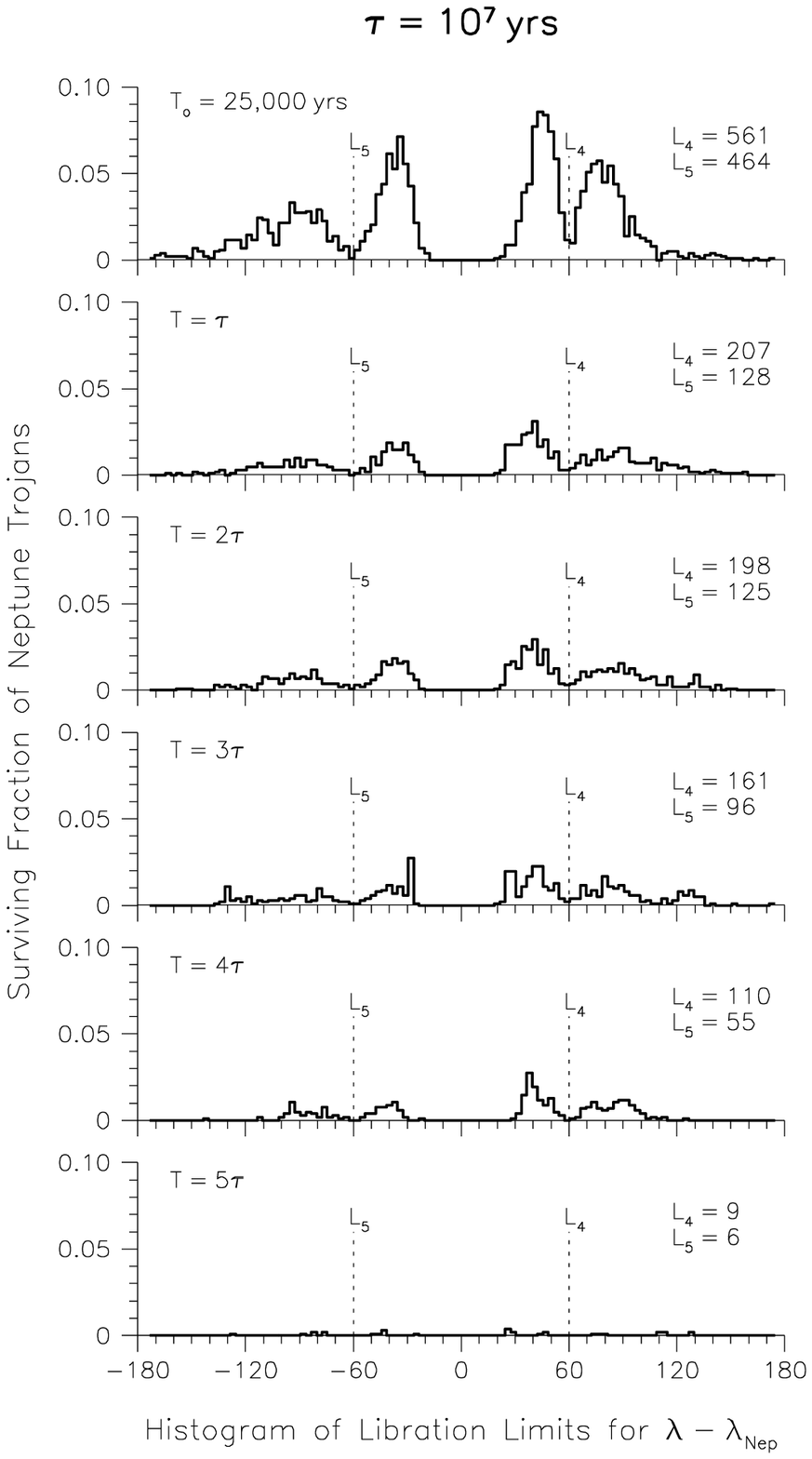}}
	\vspace{3cm}
	Figure \ref{fig:tau_7}: 
	\captauseven
\end{figure}
\clearpage
 
\begin{figure}  \refstepcounter{figs} \label{fig:tau_histogram}
	\vspace{-2cm} 
	\centerline{ 
	\hspace{-1cm}  
	\epsfxsize=6.5in  
	\epsffile[18 144 592 718]{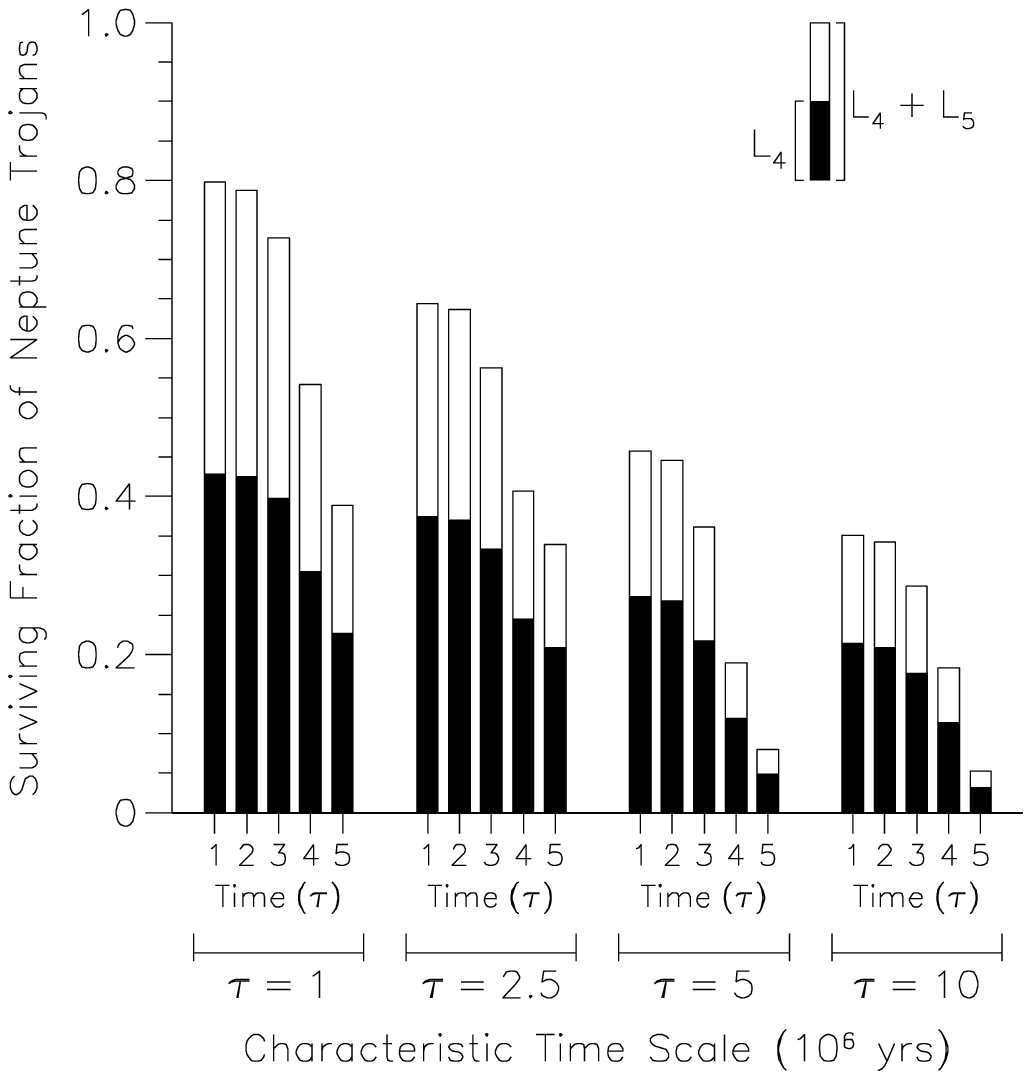}}
	\vfill
	Figure \ref{fig:tau_histogram}: 
	\captauhist
\end{figure}
\clearpage
 
\begin{figure}  \refstepcounter{figs} \label{fig:jsn}
	\vspace{-2cm} 
	\centerline{ 
	\hspace{-1cm}  
	\epsfxsize=6.5in  
	\epsffile[18 144 592 718]{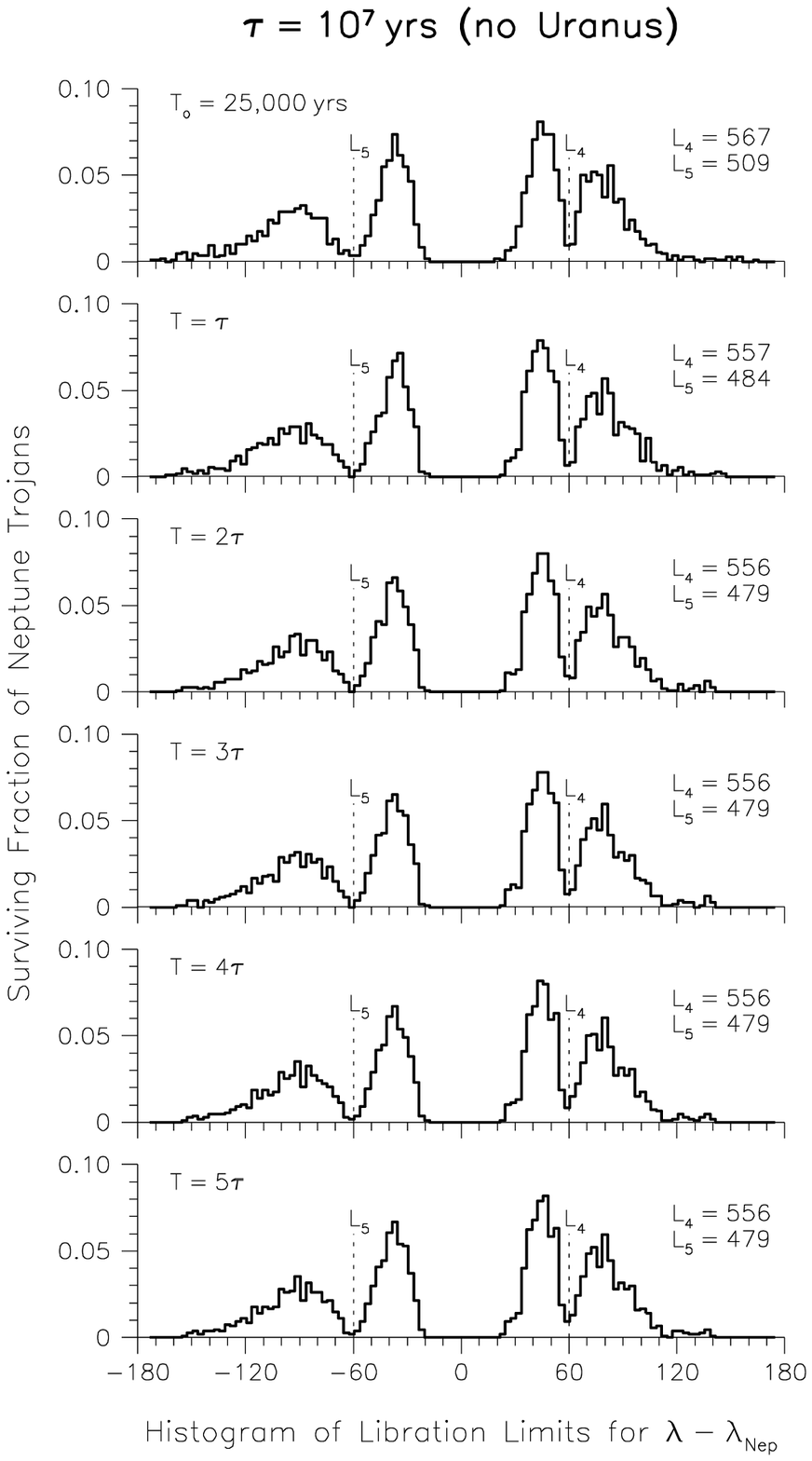}}
	\vspace{3cm}
	Figure \ref{fig:jsn}: 
	\capjsn
\end{figure}
\clearpage
 
\clearpage
 
\clearpage
 
\clearpage
 
\begin{figure}  \refstepcounter{figs} \label{fig:spectra_1tau}
	\vspace{-1cm} 
	\centerline{ 
	\hspace{-1cm}  
	\epsfxsize=6.5in  
	\epsffile[18 144 592 718]{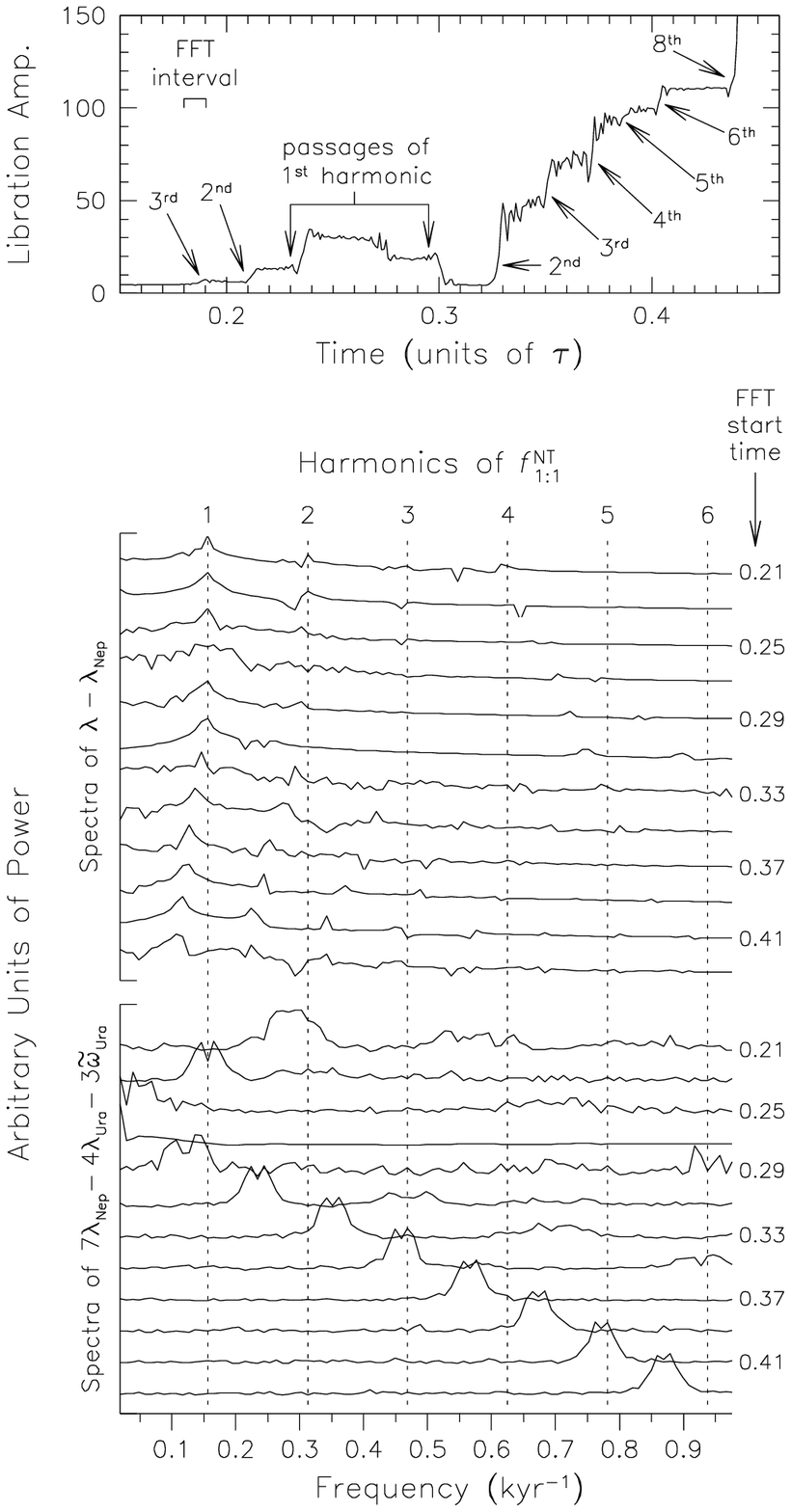}}
	\vspace{1.5cm}
	Figure \ref{fig:spectra_1tau}: 
	\capspectraonetau
\end{figure}
\clearpage
 
\begin{figure}  \refstepcounter{figs} \label{fig:final_ele}
	\vspace{-2cm} 
	\centerline{ 
	\hspace{-1cm}  
	\epsfxsize=6.5in  
	\epsffile[18 144 592 718]{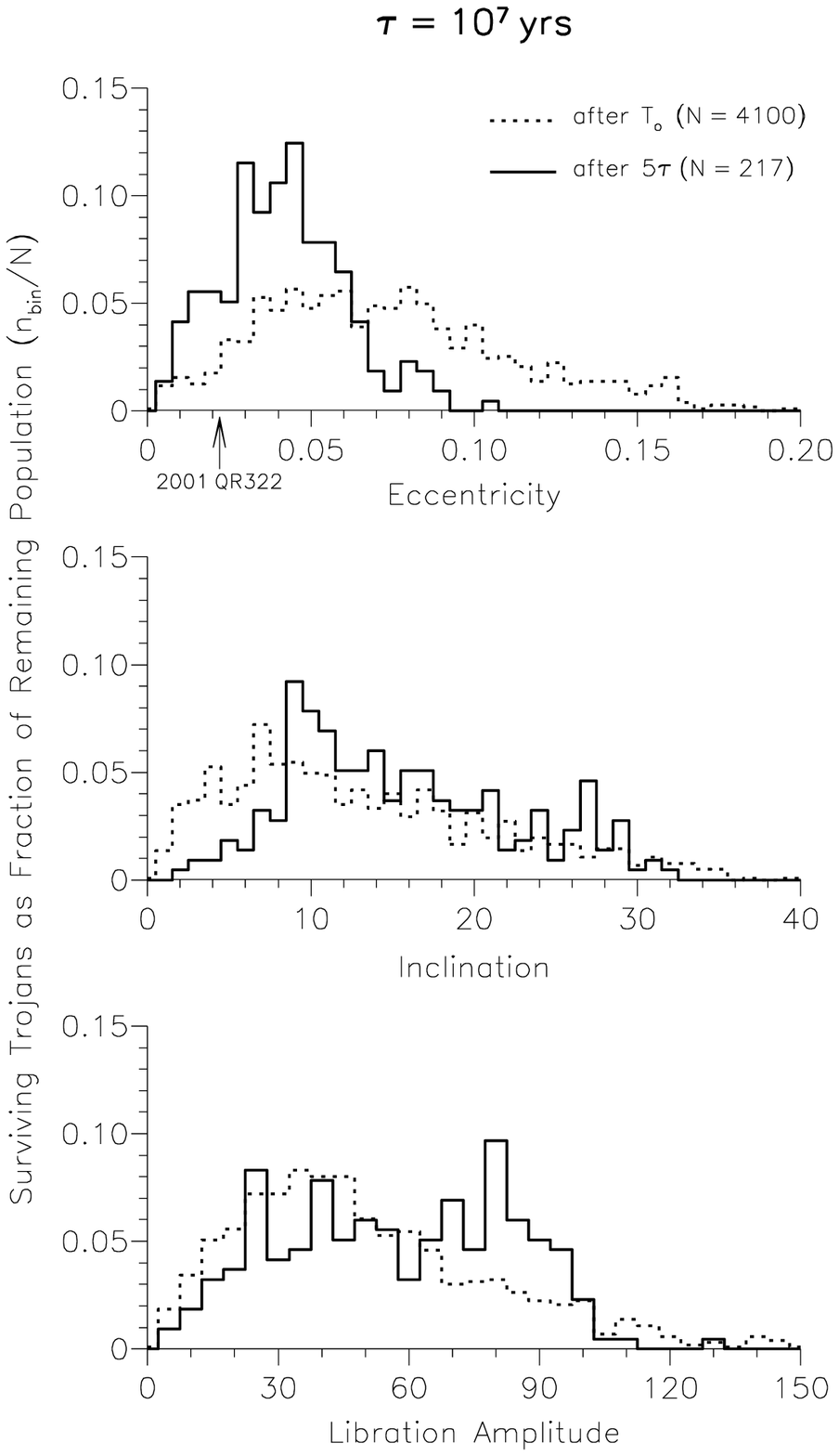}}
	\vspace{2cm}
	Figure \ref{fig:final_ele}: 
	\capfinalele
\end{figure}
\clearpage

\begin{figure}	\refstepcounter{figs} \label{fig:ecc_tau} \label{last_figure}
        \vspace{-2cm}
        \centerline{
        \hspace{-1cm}
        \epsfxsize=6.5in
        \epsffile[18 144 592 718]{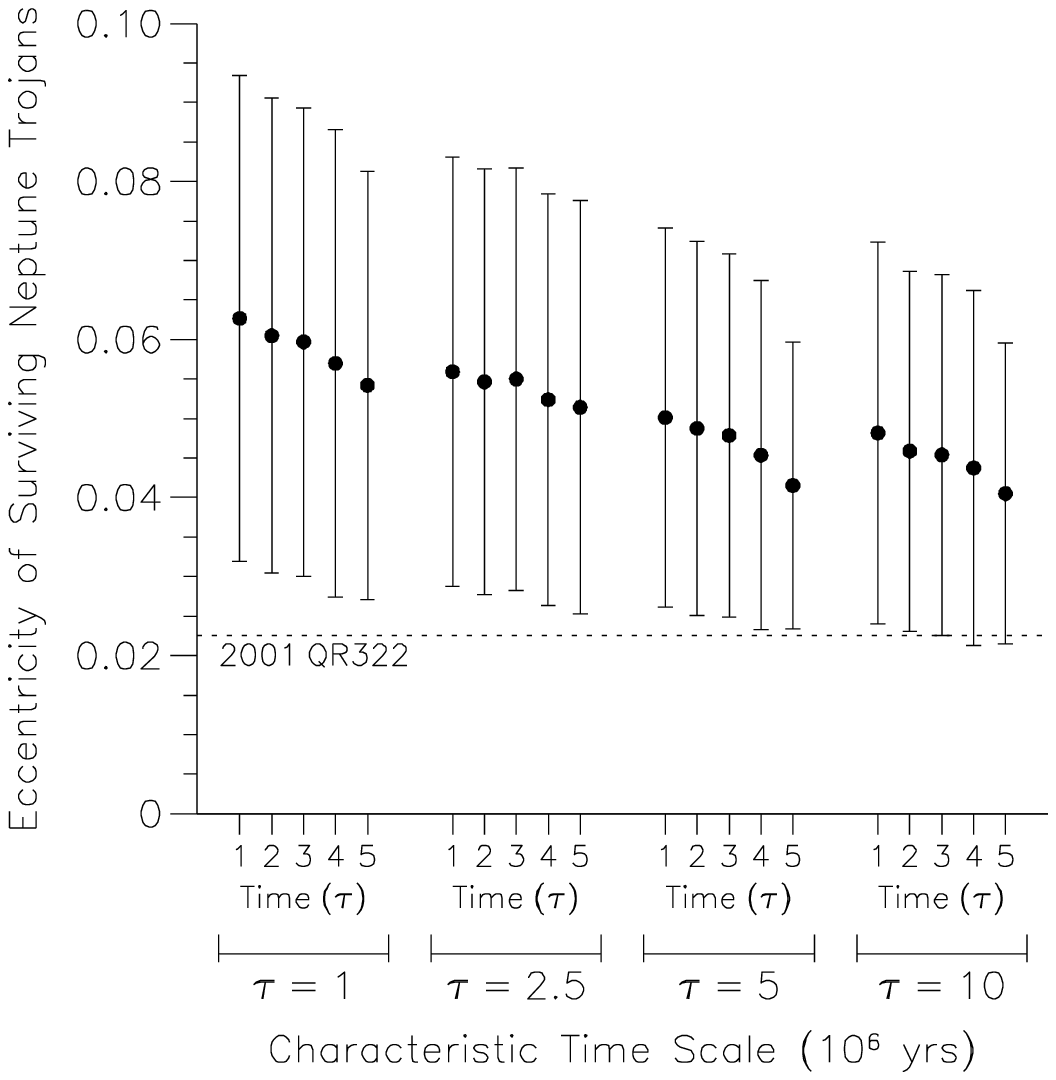}}
        \vfill
        Figure \ref{fig:ecc_tau}:
        \capecctau
        \mylabel{last_total_page}
\end{figure}

\end{document}

%% file: table1.tex
   
\renewcommand{\footnoterule}{\vspace{-0.2cm}}
\refstepcounter{tabs}
\label{tab:model_parameters}
\label{last_table}
\begin{minipage}{15cm}
\small
 \begin{tabular}{c@{\hspace{1cm}}ccc}
 \multicolumn{4}{c} {\bf Table~\ref{tab:model_parameters}} \\
 \multicolumn{4}{c} {Initial Heliocentric Planetary Configuration\footnote[2]{Solar masses, AUs, radians referred to invariable plane and mean equinox, Epoch 243000.5}} \\[0.2cm] 
 \hline \hline \\[-0.2cm] 
  Planet	& Mass \\
		& Semi-Major Axis & Eccentricity & Inclination  \\
		& Long.~Ascend.~Node & Arg.~Pericenter & Mean Anomoly  \\[0.2cm] 
 \hline \\[-0.2cm] 
  Jupiter &  9.54791810627724E$-$04 \\ 
  	  &  5.40430411233366E$+$00 & 4.90137254366321E$-$02 & 6.88824194620820E$-$03 \\
  	  &  5.45582894296211E$+$00 & 1.08695015636752E$+$00 & 5.07600450707522E$-$01 \\[0.5cm]
  Saturn  & 2.85585440033128E$-$04 \\ 
  	  & 8.78367171921144E$+$00 & 5.62633466451345E$-$02 & 1.49881886264974E$-$02 \\
  	  & 2.18380557155612E$+$00 & 5.67051860572230E$+$00 & 5.55873741721312E$+$00 \\[0.5cm]
  Uranus  & 4.37275778048887E$-$05 \\ 
  	  & 1.63160636907320E$+$01 & 4.47359382069131E$-$02 & 1.91275920013027E$-$02 \\
  	  & 5.41937420992345E$+$00 & 3.71536831241412E$+$00 & 4.48013898742567E$+$00 \\[0.5cm]
  Neptune & 5.17762233001967E$-$05 \\ 
  	  & 2.29867881794121E$+$01 & 1.18546237182779E$-$02 & 1.26187650361650E$-$02 \\
  	  & 3.49327271961024E$+$00 & 3.51610962057474E$+$00 & 2.33495547884702E$+$00 \\[0.5cm]
  \hline \hline
 \end{tabular}
\normalsize
\end{minipage}